\documentclass{article}

\usepackage[english]{babel}

\usepackage[letterpaper,top=2cm,bottom=2cm,left=3cm,right=3cm,marginparwidth=1.75cm]{geometry}

\usepackage{amsmath}
\usepackage{graphicx}
\usepackage[colorlinks=true, allcolors=blue]{hyperref}
\usepackage{bm}
\usepackage{float}
\usepackage{bbold}
\usepackage{verbatim}
\usepackage[table]{xcolor}
\definecolor{highlight}{RGB}{135,206,250} 
\usepackage{comment}
\usepackage{subfigure}
\usepackage{subcaption}
\usepackage{multirow}
\usepackage{booktabs}
\usepackage{natbib}
\usepackage{tikz}     
\usepackage{placeins}
\usepackage{rotating}

\title{Bayesian semiparametric modelling of biomarker variability in joint models}

\author{
Sida Chen\textsuperscript{1}\thanks{Corresponding author: \texttt{sida.chen@mrc-bsu.cam.ac.uk}},
Jessica K. Barrett\textsuperscript{1},
Marco Palma\textsuperscript{1,2},
Jianxin Pan\textsuperscript{3},
Brian D. M. Tom\textsuperscript{1,4}\\[0.6ex]
\parbox{\textwidth}{\centering\small
\textsuperscript{1}MRC Biostatistics Unit, University of Cambridge, Cambridge, U.K.\\
\textsuperscript{2}Great Ormond Street Institute of Child Health, University College London, London, U.K.\\
\textsuperscript{3}Guangdong Provincial Key Laboratory of Interdisciplinary Research and Application for Data Science,\\
Beijing Normal-Hong Kong Baptist University, Zhuhai, China\\
\textsuperscript{4}Yale School of Public Health, Yale University, New Haven, U.S.A.
}
}

\begin{document}
\maketitle

\begin{abstract}
There is growing interest in the role of within-individual variability (WIV) in biomarker trajectories for assessing disease risk and progression. A trajectory-based definition that has attracted recent attention characterises WIV as the curvature-based roughness of the latent biomarker trajectory (TB-WIV). To rigorously evaluate the association between TB-WIV and clinical outcomes and to perform dynamic risk prediction, joint models for longitudinal and time-to-event data (JM) are necessary. However, specifying the longitudinal trajectory is critical in this framework and poses methodological challenges. In this work, we investigate three Bayesian semiparametric approaches for longitudinal modelling and TB-WIV estimation within the JM framework to improve stability and accuracy over existing approaches. Two key methods are newly introduced: one based on Bayesian penalised splines (P-splines) and another on functional principal component analysis (FPCA). Using extensive simulation studies, we compare their performance under two important TB-WIV definitions against established approaches. Our results demonstrate overall inferential and predictive advantages of the proposed P-spline and FPCA-based approaches while also providing insights that guide method choice and interpretation of inference results. The proposed approaches are applied to data from the UK Cystic Fibrosis Registry, where, for the first time, we identify a significant positive association between lung function TB-WIV and mortality risk in patients with cystic fibrosis and demonstrate improved predictive performance for survival.
\end{abstract}

\section{Introduction}
\label{sec:intro}

In clinical and epidemiological cohort studies, biomarker data are often collected longitudinally alongside event history data on outcomes such as disease progression or mortality. One important goal is to identify informative biomarkers associated with these outcomes, so that they can be used for monitoring and predicting disease trajectories. To extract information from biomarker data, the mean level is conventionally used as a summary because of its simplicity and ease of interpretation. However, alternative characteristics may provide important complementary information, which has motivated recent studies to look beyond the mean and consider higher-order features. Within-individual variability (WIV), in particular, has attracted increasing attention as an additional measure for assessing disease risk and progression. For instance, recent clinical studies have revealed important links between blood pressure variability and a range of chronic diseases, including stroke and vascular events \citep{rothwell2010prognostic}, dementia \citep{nagai2015visit}, and diabetes \citep{bell2018prognostic}, all of which are prevalent in ageing populations and contribute substantially to disease burden. Another important clinical context that motivates this paper comes from cystic fibrosis (CF) research. CF is a genetic disorder caused by mutations in the CFTR gene, which impair salt transport and lead to mucus accumulation and progressive respiratory disease. Although relatively rare - with about 11000 individuals currently affected in the UK \citep{UKCF_Report}, CF is associated with high morbidity and reduced life expectancy. This has motivated increasing efforts to improve understanding of disease mechanisms and treatment response, and to assess the role of biomarker WIV for CF progression and mortality \citep{morgan2016forced,palma2025bayesian}.

In earlier clinical literature, naive measures of WIV, such as the standard deviation of all measurements and the coefficient of variation, were commonly used. However, these suffer from important limitations, as they are affected by subject-specific data settings (e.g. number of measurements) and by measurement error. The resulting unreliability in WIV estimates can lead to regression dilution bias when analysing their association with the outcome of interest (e.g. death), and can also undermine the predictive performance of models that include WIV as a predictor \citep{barrett2019estimating}.
More advanced and appropriate statistical methods for modelling biomarker WIV and assessing its clinical relevance have been proposed, built within a joint longitudinal and time-to-event model (JM) framework \citep{papageorgiou2019overview,wang2024joint}. In this framework, both the longitudinal process, from which WIV is derived, and the time-to-event outcome are modelled simultaneously, allowing valid inference on the association between WIV and event risk, as well as principled and dynamic risk prediction.
With regard to WIV definitions, two main classes can be distinguished, arising from different perspectives. The first, which we refer to as residual-based biomarker WIV (RB-WIV), characterises variability based on the residual dispersion around the subject-specific dominant trend \citep{gao2011joint,barrett2019estimating,courcoul2023location}. RB-WIV is typically constructed using the mixed-effects location–scale model \citep{hedeker2008application}, where separate random-effects components are specified for the subject-specific mean and for the residual variance processes. The mean process is usually modelled with a relatively simple structure to capture the long-term trend, while WIV is quantified through the subject-specific, possibly time-varying residual variance. A caveat with this framework is that RB-WIV can potentially mix biological variability with measurement error.
The second approach, which we refer to as trajectory-based biomarker WIV (TB-WIV), defines variability as fluctuations in the subject-specific latent biomarker trajectory \citep{wang2024modeling,wang2024weighted}. The construction of TB-WIV builds on a flexible model of the individual’s mean process to capture biological variability and within-subject correlation in the repeated measurements, while the residual variance represents only measurement error. WIV is then quantified through the integrated (weighted) curvature of the underlying mean function. A caveat with this framework is that TB-WIV can be sensitive to the specification of the trajectory function.
At a high level, the two WIV measures capture distinct aspects of biomarker variability, and depending on the application context, either measure may be relevant and appropriate to consider.

In this paper, we adopt the TB-WIV perspective, both because its clinical relevance remains poorly understood and because it poses important open methodological challenges. TB-WIV is defined through functionals of the second derivatives of the underlying process. As noted earlier, within this framework, specification of the subject-specific trajectories is critical. Existing implementations of TB-WIV rely on regression spline–based approaches \citep{wang2024modeling,wang2024weighted}. However, the estimation results can be sensitive to the number and location of knot points, which is unsurprising given that knot configuration directly governs the flexibility of the resulting trajectories. Although empirical strategies, such as information criteria, can guide knot selection, choice of a suitable configuration is not guaranteed.
In fact, derivative estimation in longitudinal settings has received very little attention. To our knowledge, \citet{simpkin2018derivative} is the only relevant work that investigates the representation and estimation of derivative processes for longitudinal data. Their analysis was restricted to first- and second-order derivatives, and although they explored regression splines and functional principal component analysis (FPCA), the potential as well as limitations of such approaches was not fully exploited, with the scope limited to the trajectory process alone.

In this work, we explore three Bayesian semiparametric approaches for longitudinal modelling and TB-WIV estimation within the JM context, all of which bypass strong dependence on knot settings. We propose two new approaches that build on state-of-the-art penalised splines (P-splines) and sparse FPCA, offering generic modelling flexibility. We also consider the semiparametric multiplicative random-effects model (SMRE), proposed by \citet{luo2025semiparametric} for parsimonious modelling of TB-WIV, which we adapt to the Bayesian JM setting.
We adopt a Bayesian framework because it allows regularisation through priors and facilitates the handling of potentially large numbers of random effects via efficient Markov chain Monte Carlo (MCMC) sampling. Using extensive simulation studies, we empirically compare these approaches under both hypothetical and realistic settings. Our results provide important practical insights into their performance for inference on key model parameters, including TB-WIV associations, as well as implications for survival prediction.
Motivated by a recent analysis of RB-WIV using the UK CF Registry \citep{palma2025bayesian}, we analysed the same database and investigated, for the first time, the association between TB-WIV of forced expiratory volume in one second (FEV1), a lung function biomarker, and the risk of death in patients with CF. Our findings offer clinical insights into the prognostic relevance and added predictive value of lung function TB-WIV for CF mortality.

The rest of the paper is organised as follows. We introduce the joint modelling framework with TB-WIV in Section~\ref{sec:JM-WIV}. Section~\ref{sec:methods-WIV} presents three Bayesian semiparametric approaches for modelling the longitudinal biomarker trajectory and TB-WIV, and Section~\ref{sec:sim} describes the design and results of the simulation studies. The proposed approaches are then applied to the UK CF Registry data in Section~\ref{sec:app}, and the paper concludes with discussion and future perspectives in Section~\ref{sec:disc}.

\section{Joint models incorporating TB-WIV}
\label{sec:JM-WIV}

\subsection{Model formulation}
\label{subsec:JM-WIV-formulation}
Let $y_{ij}=y_i(t_{ij})$ be the $j$-th longitudinal measurement of a biomarker collected from the $i$-th subject, recorded at time points $t_{ij}$, where $i=1,\ldots,n$ and $j=1,\ldots,n_i$. Considering a generic longitudinal model for $y_i(t)$
\begin{equation}\label{eq:jm-wiv-long}
    y_i(t)=w_{L,i}^{\top}\beta_L+\mu_i(t)+\epsilon_i(t), \quad \epsilon_i(t)\sim N(0,\sigma_{e}^{2}),
\end{equation}
where $w_{L,i}$ is a vector of baseline covariates related to the biomarker trajectory and $\beta_L$ is the associated coefficient vector, $\mu_i(t)$ represents the underlying, subject-specific trajectory of the biomarker and $\epsilon_i(t)$ represents the random measurement error. 
TB-WIV characterises the variability pattern arising from the inherent biological fluctuation in the underlying trajectory $\mu_i(t)$ in equation \eqref{eq:jm-wiv-long}, separating from the variability contributed by measurement error. The seminal proposal in \citet{wang2024modeling} is motivated by smoothing spline roughness penalties and defines TB-WIV as the cumulative roughness of the individual's latent trajectory
\begin{equation}\label{eq:TB-WIV-cumcurv}
    \int_{t_0}^t \{ \mu_i''(s) \}^2 \, ds,
\end{equation}
where $t_0$ is a pre-specified initial time for computing TB-WIV (e.g. to exclude an early post-treatment phase if needed). In this paper, we set $t_0=0$ throughout. In \citet{wang2024weighted}, this measure was extended to a weighted cumulative form
\begin{equation}\label{eq:TB-WIV-weightedcumcurv}
    \int_{t_0}^t \omega_{\sigma}(t - s) \{ \mu_i''(s) \}^2 \, ds,
\end{equation}
where $\omega_{\sigma}(t - s)$ is a weighting function (e.g., a truncated Gaussian density) with scale parameter $\sigma$ that controls the down-weighting of past variability contributions.
 
In addition to longitudinal data, survival data for an event of interest are also recorded for each individual. In the JM framework, the time-to-event process is modelled simultaneously, allowing us to examine how certain characteristics of the longitudinal trajectory relate to the event process. In our context, we are particularly interested in understanding the association and incremental predictive value of TB-WIV for the risk of the event of interest.
Let $T^{*}_i$ and $C_i$ denote the true event time and the right-censoring time, respectively. The observed follow-up time is $T_i=\min \{T^{*}_i, C_i\}$ and $\delta_i=I(T^{*}_i\leq C_i)$ is the event indicator, where $I(\cdot)$ is the indicator function. To specify the survival submodel, we consider a relative hazard model, which is the most common choice in such joint modelling contexts. The hazard function $h_i(t)=\lim_{\Delta t\to 0}\frac{1}{\Delta t}P(t\leq T^{*}_{i}<t+\Delta t \mid T^{*}_{i}\geq t)$ is modelled as
\begin{equation}\label{eq:jm-wiv-surv}
h_i(t)=h_0(t)\exp\big(w_{S,i}^{\top}\gamma+\alpha_1 m_i(t)+\alpha_2\text{WIV}_i(t)\big),
\end{equation}
where $h_0(t)$ is the baseline hazard function, $w_{S,i}$ is the vector of baseline covariates associated with the event process, and $\gamma$ is the corresponding coefficient vector. 
The term $m_i(t)=w_{L,i}^\top\beta_L+\mu_i(t)$ denotes the subject-specific underlying biomarker value, and $\text{WIV}_i(t)$ denotes the measure of TB-WIV, both at time $t$.
In this paper, we focus on two forms of $\text{WIV}_i(t)$: the square root of the unweighted cumulative version in \eqref{eq:TB-WIV-cumcurv}, $\sqrt{\int_{0}^{t} \{\mu_i''(s)\}^2\,ds}$, and the instantaneous version $\sqrt{\{\mu_i''(t)\}^2}=|\mu_i''(t)|$.
The former captures the accumulation of curvature of the trajectory up to $t$, whereas the latter captures the local curvature at time $t$ and is, to first order, approximately related to the one-year window measure $\sqrt{\int_{t-1}^{t}\{\mu_i''(s)\}^2\,ds}$.
Note that we adopt the square-root transform when defining $\text{WIV}_i(t)$, following \citet{wang2024modeling,wang2024weighted}, so that this covariate is on the same order of magnitude as the biomarker level $m_i(t)$.
We do not consider the weighted cumulative measure in  \eqref{eq:TB-WIV-weightedcumcurv} for two reasons. First, as noted in \citet{wang2024weighted}, estimation can suffer from weak identifiability when the data contain limited information (e.g., when the number of measurements per subject is small). Second, the two specific TB-WIV definitions we consider can be viewed as important special cases of the weighted cumulative measure, corresponding respectively to a uniform weighting function or a Dirac delta function at the current time. 
They are of methodological interest in their own right, as shown later in the simulation study, where they exhibit distinct impacts on the estimation performance of the modelling approaches considered.
$\alpha_1$ and $\alpha_2$ are coefficients characterising the associations of, respectively, the current biomarker level and $\text{WIV}_i(t)$, with the risk of the event. In particular, $\alpha_2>0$ indicates that, with other covariates held fixed, higher biomarker variability (as defined via $\text{WIV}_i(t)$) is associated with greater instantaneous risk of the event at time t, and vice versa.
In this paper, motivated by our application context, we restrict our attention to the formulation in \eqref{eq:jm-wiv-surv}. Other association structures could also be considered; for example, the hazard may alternatively depend on a (weighted) cumulative version of the trajectory $m_i(t)$ or on other potentially relevant features.

\subsection{The inferential challenge}
Once the longitudinal and survival submodels are specified, estimation of the joint model reduces to a standard problem, commonly addressed either by maximising a semiparametric joint likelihood after integrating out all latent variables (see \citealp{wang2024modeling,wang2024weighted}) or by adopting an MCMC-based, fully Bayesian approach, as considered in this paper. The challenges, however, arise from the model specification for $\mu_i(t)$, as this choice impacts estimation of the underlying trajectory and the TB-WIV measure $\text{WIV}_i(t)$, and consequently the association parameters $\alpha_1$ and $\alpha_2$, and may even influence other model parameters.

The TB-WIV measures are particularly challenging to estimate, as they depend on higher-order features of the trajectory function $\mu_i(t)$, which are inherently noisier, and small errors in $\mu_i(t)$ may be magnified when propagated through derivatives or curvature functionals. Existing approaches proposed by \citet{wang2024modeling,wang2024weighted} rely on regression splines as a sieve approximation to the longitudinal trajectory. However, our experimental results show that misspecifying the knot configuration can lead to substantial bias in parameter estimates (see Section \ref{sec:sim}). This highlights the need for further exploration of this issue, as the sensitivity poses concerns regarding the reliability of these approaches.

\section{Bayesian semiparametric modelling approaches for characterising TB-WIV}
\label{sec:methods-WIV}

In this section, we introduce three Bayesian semiparametric approaches for modelling the subject-specific biomarker trajectory and the associated TB-WIV within the joint modelling framework of Section \ref{sec:JM-WIV}.

\subsection{The P-spline-based approach}
\label{subsec:methods-WIV-pspline}
B-splines with random effect weights have been used as a standard approach in the JM context for flexible trajectory modelling, and have also been adopted for modelling TB-WIV. Specifically, the subject-specific longitudinal trajectory is commonly modelled as
\begin{equation}\label{eq:rspline}
    \mu_i(t)=\sum_{k=1}^{K}(\beta_k+b_{ik})B_k(t),
\end{equation}
where $B_k(t)$ is the k-th cubic B-spline basis function defined on pre-specified knot points with boundary knots covering observation domain $\mathcal{T}$, $\beta=(\beta_1,\ldots,\beta_K)^\top$ denotes fixed effects and $b_i=(b_{i1},\ldots,b_{iK})^\top\sim N(0,\Sigma_b)$ denotes subject-specific random effects, with $\Sigma_b$ the random effects covariance matrix. Under this formulation, TB-WIV can be expressed as a quadratic form in the spline coefficient vector $\beta+b_i$. However, a key modelling challenge with this approach is knot placement. This motivates our consideration of a Bayesian P-spline–based approach. Bayesian P-splines were originally introduced by \citet{lang2004bayesian} as a Bayesian formulation of the frequentist P-spline smoothing framework of \citet{eilers1996flexible}, for estimating smooth functions within structured additive regression. The basic idea is that, rather than relying on careful knot choice, a large number of basis functions is employed to allow sufficient flexibility, while regularising priors on the spline coefficients (e.g., a second-order random walk (RW2) prior) are used to discourage overfitting. The P-spline construction thus greatly reduces sensitivity to knot placement compared with regression splines. In the longitudinal setting, however, a naive implementation can be computationally challenging: the potentially large number of random effects leads to a highly complex parameter space, and strong collinearity among the raw basis functions can further hinder MCMC mixing and slow convergence.

To address this, we adapt the orthogonalised P-spline construction of \citet{scheipl2012spike}, originally developed for structured additive regression, to our longitudinal framework. Specifically, we represent the subject-specific trajectory as 
\begin{equation}\label{eq:pspline}
    \mu_i(t)=\bar{\mu}(t)+b_i(t), \quad b_i(t) = b_{i0}+b_{i1}t+\sum_{k=1}^{K}\zeta_{ik}\tilde{B}_k(t),
\end{equation}
where $\bar{\mu}(t)$ is the population mean function shared across subjects, and $b_i(t)$ characterises subject-specific deviations. 
For $\bar{\mu}(t)$, we approximate it using standard Bayesian P-splines \citep{lang2004bayesian}, which work well for estimating a single smooth curve. More specifically, we set $\bar{\mu}(t)=\sum_{j=1}^{M}\beta_jB_j(t)$, with a RW2 prior on the spline coefficients $\beta$, which yields a structured multivariate normal prior $\beta\sim N(0,(\tau_\beta \tilde{P})^{-1})$, where $\tilde{P}=D_{2,M}^\top D_{2,M}+10^{-6} I_M$. Here $D_{2,M}$ is the second-order difference matrix of dimension $(M-2)\times M$ and $I_M$ is the identity matrix of dimension M. The small ridge term $10^{-6} I_M$ is added to ensure $\tilde{P}$ is full rank for numerical stability. Following standard practice, the smoothing precision parameter $\tau_\beta$ is given a $\text{Gamma}(0.01,0.01)$ prior.
For $b_i(t)$, we orthogonalise the spline basis, splitting the spline space into the unpenalised null space (spanned by constant and linear terms) and its penalised orthogonal complement (containing no linear component). The $\tilde{B}_k(t)$ are orthogonalised basis functions obtained by performing a spectral decomposition of $BP^{-}B^\top$, where B is the original B-spline design matrix (formed on a dense grid within the time domain) with $K_0$ basis functions and $P^{-}$ denotes the generalised inverse of the usual second-order difference penalty matrix $P=D_{2,K_0}^\top D_{2,K_0}$, where $D_{2,K_0}$ is the second-order difference matrix of dimension $(K_0-2)\times K_0$. The subject-specific random coefficients $b_{i0}$ and $b_{i1}$ correspond to the null space, while $\zeta_{ik}$ parameterise the penalised part. In practice, only eigenvectors of the penalised part that explain all practically relevant variation (e.g. $99.9\%$) are retained, and the effective dimension $K$ is typically far smaller than the original basis dimension $K_0$ (e.g., in our implementation we used $K_0=40$; with our $99.9\%$ retention rule this yielded $K=10$). 
Another practical advantage of the orthogonal P-spline representation in \eqref{eq:pspline} is that it greatly simplifies prior specification and facilitates MCMC sampling. With the raw B-spline basis, one typically imposes a random walk type prior directly on the subject-specific spline coefficients, which induces strong dependence among random effects and a complex posterior geometry. By contrast, the orthogonalised basis is decorrelated, and therefore independent priors can be placed on the associated coefficients. To flexibly adapt to subject-specific variability while allowing information borrowing within and across subjects, we consider a weakly informative, local-global type prior of the form
\begin{equation}
    \zeta_{ik}\sim N(0,(s_i\tau_k)^2), \quad s_i\sim \text{Half-N}(0,5^2), \quad \tau_k\sim \text{Gamma}(2,1),
\end{equation}
where $\text{Half-N}(0,\sigma^2)$ denotes a half-normal distribution on the positive range with scale parameter $\sigma$ and $\text{Gamma}(a,b)$ denotes a Gamma distribution with shape $a$ and rate $b$. Other reasonable priors can be used for $s_i$ and $\tau_l$, and, in our experiments, results are generally robust. For the random intercept and slope we assign independent Gaussian priors $b_{i0}\sim N(0,\sigma_{b0}^2)$ and $b_{i1}\sim N(0,\sigma_{b1}^2)$, with variance parameters given a vague $\text{Inverse-Gamma}(0.01,0.01)$ hyperprior, where $\text{Inverse-Gamma}(a,b)$ denotes a inverse Gamma distribution with shape $a$ and scale $b$.
With the model defined above, it follows that 
\begin{equation}
\begin{aligned}\label{eq:p-spline-TBWIV}
\mu_i''(t)
  &= \sum_{l=1}^{M} \beta_l\, B_l''(t)
   + \sum_{k=1}^{K} \zeta_{ik}\, \tilde B_k''(t), \\
\int_{0}^{t} \{\mu_i''(s)\}^2 \, ds
  &= \beta^\top P_{B}(0,t)\, \beta
   + 2\, \beta^\top P_{B,\tilde B}(0,t)\, \zeta_i
   + \zeta_i^\top P_{\tilde B}(0,t)\, \zeta_i ,
\end{aligned}
\end{equation}
where $\beta=(\beta_1,\ldots,\beta_M)^\top$, $\zeta_i=(\zeta_{i1},\ldots,\zeta_{iK})^\top$, $P_{B}(0,t)\in\mathbb{R}^{M\times M}$, $P_{B,\tilde B}(0,t)\in\mathbb{R}^{M\times K}$, and $P_{\tilde B}(0,t)\in\mathbb{R}^{K\times K}$, with entries $[P_{B}(0,t)]_{lm}=\int_{0}^{t} B_l''(s) B_m''(s)\,ds$, $[P_{B,\tilde B}(0,t)]_{lk}=\int_{0}^{t} B_l''(s)\,\tilde B_k''(s)\,ds$, and $[P_{\tilde B}(0,t)]_{kk'}\\
=\int_{0}^{t} \tilde B_k''(s)\,\tilde B_{k'}''(s)\,ds$.

\subsection{The FPCA-based approach}
\label{subsec:methods-WIV-fpca}
Suppose the subject-specific trajectories $\mu_i(t)$ are independent realisations of a smooth, square-integrable random function $\mu(t)$ on domain $\mathcal{T}$, and assume the existence of a continuous mean function $\bar{\mu}(t)=E[\mu(t)]$ and covariance kernel $\sigma(s,t)=E[(\mu(s)-\bar{\mu}(s))(\mu(t)-\bar{\mu}(t))]$. 
FPCA builds on the Karhunen–Lo\`eve (KL) expansion for functional data (i.e., time-indexed curves/trajectories; see \citealp{ramsay2005functional}), which represents each trajectory using an infinite basis expansion
\begin{equation}\label{eq:fpca-KL}
    \mu_i(t)=\bar{\mu}(t)+\sum_{l=1}^{\infty}\zeta_{il}\psi_l(t),
\end{equation}
where $\{\psi_l(t)\}$ are orthonormal eigenfunctions of the covariance kernel $\sigma(s,t)$, defined by $\int_{\mathcal{T}}\sigma(s,t)\psi_l(s)ds\\
=\lambda_l\psi_l(t)$, with eigenvalues $\lambda_1\geq\lambda_2\geq\lambda_3\cdots\geq0$. The $\zeta_{il}$ are uncorrelated random variables with mean zero and variance $\lambda_l$, and their variances $\lambda_l$ characterise the amount of variation in the direction of $\psi_l(t)$.
FPCA is the functional extension of PCA, providing a flexible yet parsimonious representation of sparse and irregular longitudinal trajectories, as often encountered in clinical settings, using only a small number of components (eigenfunctions) \citep{yao2005functional}.
Notably, FPCA achieves optimal dimension reduction: for any positive integer L, the truncated KL expansion obtained using the first L eigenfunctions minimises the expected integrated squared error among all L-dimensional basis representations of the centred process $\mu_i(t)-\bar{\mu}(t)$, and the minimal error is $\sum_{l>L}\lambda_l$ \citep{ramsay2005functional}. This optimality provides an efficiency advantage over the spline-based approach.
The truncation level L therefore controls the approximation accuracy. In practice, L is often chosen to achieve a high proportion of variance explained (PVE), i.e., $\sum_{l=1}^{L}\lambda_l/\sum_{k\geq1}\lambda_k$, balancing accuracy and parsimony. The retained eigenfunctions $\{\psi_1(t),\ldots,\psi_L(t)\}$ summarise the dominant modes of variation in the centred trajectories, where larger eigenvalues $\lambda_l$ indicate a greater contribution of the $l$-th mode to the total functional variance.

In our Bayesian implementation, we estimate the eigenfunctions in a preliminary step based on covariance smoothing of the observed longitudinal data, which can alternatively be viewed as providing data-adaptive basis functions. Estimating them jointly with other parameters in the JM would greatly increase computational complexity, with negligible gains since the survival data provide little information on the eigenfunctions. For this purpose, we employ the sparse FPCA techniques of \citet{xiao2018fast}, implemented in the R package FACE \citep{face}. It directly estimates the covariance operator from sparse, irregular data via P-spline smoothing with cross-validation–based smoothness tuning, delivering accurate eigenfunction estimation with strong computational efficiency, and is state of the art for sparse FPCA. To complete the Bayesian specification, the associated subject-specific coefficients $\zeta_{il}$ (interpreted as FPC scores in standard FPCA) are assigned independent Gaussian priors $\zeta_{il}\sim N(0,\nu_l^{2})$ for $i=1,\ldots,n$ and $l=1,\ldots,L$. For greater data-adaptiveness and to avoid over-reliance on the preliminary FPCA eigenvalues, each variance parameter $\nu_l^{2}$ is treated as a free parameter rather than being fixed to its FPCA eigenvalue $\lambda_l$ (as in the classical KL formulation), and is given a vague $\text{Inverse-Gamma}(0.01,0.01)$ hyperprior.
To determine the truncation level L, we use a PVE of $99.9\%$. Our experiments indicate that such a high level is desirable in the curvature estimation context, as even components in the tail may contain non-ignorable curvature information (see Section \ref{subsec:sim_disc} for additional discussion). The mean function in the KL expansion \eqref{eq:fpca-KL} is approximated using a standard Bayesian P-spline, as in the P-spline-based approach. The knot locations are chosen consistently with those used in the FACE package when estimating the eigenfunctions, to ensure coherence between the two steps.
With our FPCA parameterisation above, the expressions for $\mu_i''(t)$ and $\int_{0}^{t}\{\mu_i''(s)\}^2ds$ take the same form as in \eqref{eq:p-spline-TBWIV} with the orthogonal spline basis $\tilde B_k$ replaced by the estimated eigenfunctions $\psi_\ell$; the corresponding matrices $P_{B,\psi}(0,t)$ and $P_{\psi}(0,t)$ are defined analogously.

\subsection{The SMRE-based approach}
\label{subsec:methods-WIV-smre}
The original form of the SMRE model was proposed by \citet{luo2025semiparametric} as a simpler alternative to the regression spline-based formulations in \eqref{eq:rspline}. In this approach, individual trajectories are represented as
\begin{equation}\label{eq:SMRE}
\mu_i(t)=\beta_0+b_{i0}+(\beta_1+b_{i1})t+b_{i2}\mu(t),
\end{equation}
where $\beta_0$ and $\beta_1$ are the population-level intercept and slope, $b_{i0}$, $b_{i1}$ and $b_{i2}$ are subject-specific random effects, and $\mu(t)$ is a common smooth nonlinear function. For identification, the multiplicative random effect $b_{i2}$ is constrained to have mean one, and the condition $\lim_{t\to 0^{+}}\lvert \mu(t)\rvert/t=0$ is further imposed to remove any linear contribution from $\mu(t)$ and to ensure that the trajectory is governed by the linear term near the origin. The rationale for this SMRE specification is informed by empirical findings from \citet{luo2025semiparametric} in their application: at the population level, biomarker progression is approximately linear close to the origin, with nonlinearity emerging later.

The shape constraint imposed in the nonlinear term $\mu(t)$, however, can be overly restrictive. Individual trajectories may be highly heterogeneous and need not follow the same population-level pattern, and the early-linear/late-nonlinear structure assumed in \citet{luo2025semiparametric} may not hold in more general data settings. Here, we extend the original SMRE construction within our Bayesian modelling framework by relaxing this shape condition to achieve greater flexibility. Specifically, we retain the SMRE trajectory model in \eqref{eq:SMRE}, but model the nonlinear component $\mu(t)$ without shape constraints via Bayesian P-splines, as $\mu(t)=\sum_{j=1}^{M}\beta_jB_j(t)$, $\beta\sim N(0,(\tau_\beta \tilde{P})^{-1})$, with the basis functions $B_j(t)$, penalty matrix $\tilde{P}$, and smoothing parameter $\tau_\beta$ defined as in Section \ref{subsec:methods-WIV-pspline}. 
To anchor the scale of $\mu(t)$, we impose a Bayesian analogue of the moment constraint of \citet{luo2025semiparametric} by re-centring the subject-specific multiplicative random effects $b_{i2}$ at each MCMC iteration so that $n^{-1}\sum_{i=1}^n b_{i2}=1$.
Importantly, although we relax the shape restriction on $\mu(t)$, the resulting subject-specific trajectory $\mu_i(t)$, which is the object of inference, remains identifiable. 
From a structural perspective, our Bayesian SMRE imposes a rank-1 functional constraint on between-subject nonlinear variation. In particular, SMRE constrains all nonlinear deviations to lie in the one-dimensional subspace spanned by the common function $\mu(t)$; and $\text{Cov}(b_{i2}\mu(t),b_{i2}\mu(s))=\text{Var}(b_{i2})\mu(s)\mu(t)$, which has rank one. In this sense, SMRE provides a parsimonious, reduced rank surrogate to the more flexible but higher-dimensional P-spline approach introduced in Section \ref{subsec:methods-WIV-pspline}, for which the covariance operator of the subject-specific nonlinear deviation has rank K, where K is the number of orthogonal basis functions. 
\citet{luo2025semiparametric} proposed FPCA-based checks for assessing the appropriateness of the SMRE model, but it should be noted that these are sufficient but not necessary. When they hold, the SMRE model is well supported; when they do not, it may still be appropriate provided the rank-1 restriction gives a sufficiently accurate approximation to the subject-specific nonlinear deviation.
With the SMRE model defined above, it follows that 
\begin{equation}
\begin{aligned}
\mu_i''(t)
  &= b_{i2}\,\sum_{l=1}^{M} \beta_l B_l''(t), \\
\int_{0}^{t} \{\mu_i''(s)\}^2 \, ds
  &= b_{i2}^2\, \beta^\top P_{B}(0,t)\beta,
\end{aligned}
\end{equation}
where $\beta=(\beta_1,\ldots,\beta_M)^\top$, and $P_{B}(0,t)\in\mathbb{R}^{M\times M}$, with entries $[P_{B}(0,t)]_{lm}=\int_{0}^{t} B_l''(s)\,B_m''(s)\,ds$.

\section{Simulation study}
\label{sec:sim}

We considered several simulation cases to empirically investigate and compare the performance of three longitudinal modelling approaches described in the previous section, namely the P-spline, FPCA, and SMRE-based approaches, within the JM setup as defined in Section \ref{subsec:JM-WIV-formulation}.  As a side comparison, we also considered the regression spline (R-spline) approach with two strategies for knot placement, motivated by \citet{wang2024modeling}: using the midpoint of the follow-up period (R-spline-1), and using three equidistant quantiles of the pooled observed visit times $t_{ij}$ (R-spline-2). In each case we consider two versions of TB-WIV measures for specifying the survival submodel in equation \eqref{eq:jm-wiv-surv}: either the current curvature $\text{WIV}_i(t)=\lvert \mu_i''(t)\rvert$ or the cumulative curvature $\text{WIV}_i(t)=\sqrt{ \int_{0}^t \left( \mu_i''(s) \right)^2 \, ds}$. The key evaluation criteria were (i) the estimation accuracy of the main model parameters of interest and (ii) predictive performance, focusing on survival prediction conditional on longitudinal history, as event-risk prediction is typically the primary goal. Note that in all our simulation scenarios the modelling approaches are operated under misspecification of the longitudinal trajectory, but the TB-WIV definition used for fitting matches the data-generating definition (current vs cumulative). This is intentional to evaluate each approach as an approximation or working model. Results from the true model were obtained to serve as a gold-standard benchmark.

In our Bayesian implementation, we specified the same weakly informative priors for parameters common to all approaches. We used $N(0,10^2)$ priors for the fixed effects and association parameters, and $\text{Inverse-Gamma}(0.01,0.01)$ priors for variance parameters (measurement error and random-effects variances). For the Weibull baseline hazard (Cases 1 and 2), we used a half-Cauchy prior with scale parameter 1 for the shape parameter and an $N(0,10^2)$ prior for the log-scale parameter. For the spline-based baseline hazard (Case 3), we specified $N(0,10^2)$ priors for the spline coefficients. Priors for approach-specific parameters are as described in Section \ref{sec:methods-WIV}. We set $M=13$ spline basis functions for the mean function $\bar{\mu}(t)$ in the P-spline approach, and the same $M=13$ functions for the nonlinear component $\mu(t)$ in the SMRE approach. Experiments under our simulation settings indicated that the results were generally robust to the choice of $M$, provided it was chosen sufficiently large (e.g., $M>10$). For efficient MCMC sampling, we used the No-U-Turn Sampler (NUTS) implemented in the \texttt{rstan} R package \citep{rstan}. We adopted the default settings for the optional tuning parameters in NUTS, and based inference on 1000 iterations after discarding the first 1000 iterations as burn-in. Convergence diagnostics indicated that these choices were sufficient. 
To evaluate predictive performance, we used the leave-one-out information criterion (LOOIC; \citealp{vehtari2017practical}), which estimates out-of-sample predictive accuracy from pointwise (per-subject) log-likelihood contributions and is widely used for Bayesian JM. Because our target is survival prediction conditional on longitudinal history, the pointwise terms were defined as the conditional survival log-likelihood. Using the full joint likelihood would make the score largely driven by the longitudinal submodel (as there may be many observations per subject), potentially obscuring differences relevant for event-risk prediction.
Across all approaches, survival-based LOOIC was estimated using the \texttt{loo} R package \citep{loo}, and the Pareto-$k$ diagnostics were satisfactory, indicating reliability of the LOOIC estimates. All computations were performed on the Cambridge Service for Data Driven Discovery high-performance computing system using Ice Lake CPUs.

\subsection{Case 1: model with spline-based trajectories}

This setting follows Case 1 of the simulation study in \citet{wang2024modeling}. The longitudinal trajectories were generated using regression splines with three interior knots equally spaced on $[0,10]$, with coefficient vector $\beta=(6,\,3,\,7,\,1,\,8,\,5,\,4)$ and subject-specific coefficients $b_i\sim N\!\big(0,\mathrm{Diag}(3,4,4,5,4,3,4)\big)$. Observed data were generated on a regular visit schedule with a gap of 0.5, and measurement error variance $\sigma_e^2=1$. No covariates are included in the longitudinal process. The structure of the survival submodel was as in \eqref{eq:jm-wiv-surv}, with a single artificial covariate $w_{S,i}$ generated from a Bernoulli distribution with probability 0.5 of being 1 and associated coefficient $\gamma=-2$, and with association parameters $\alpha_1=0.2$ and $\alpha_2=0.3$. For the baseline hazard function, however, unlike \citet{wang2024modeling}, where the baseline hazard was positive and constant only after the midpoint of follow-up, we adopted a more generic Weibull baseline hazard to more realistically reflect the survival patterns observed in our real data. We set the Weibull shape parameter to 3 for both TB-WIV formulations, and the log-scale parameter to $-7$ and $-8$ for the current and cumulative curvature cases, respectively. 
Right-censoring times were generated from $\mathrm{Unif}(6,15)$, and follow-up was administratively censored at 10. Under these settings, the number of longitudinal measurements $n_i$ (truncated at each subject’s observed follow-up time) ranged from 1 to 21, with a censoring rate of approximately 40\%. We fixed the sample size at $n=1000$.

Table \ref{tab:sim_case1} compares estimation performance of the different modelling approaches in terms of bias (posterior means) and coverage probability (CP; $95\%$ credible intervals) under two types of TB-WIV associations. With the current curvature association, the P-spline- and FPCA-based approaches achieved reasonable performance in estimating the association parameters $\alpha_1$ and $\alpha_2$, with P-splines giving slightly better accuracy for $\alpha_2$. Both approaches clearly outperformed SMRE and the R-spline approaches. In particular, under both R-spline settings the estimation of $\alpha_2$ suffered from serious bias and poor coverage. All approaches estimated the fixed effect $\gamma$ well; however, only FPCA was able to recover the measurement error variance $\sigma_e^2$ well, while SMRE and R-spline-1 showed the largest bias. With the cumulative curvature association, all approaches suffered from varying levels of bias in estimating $\alpha_2$, with R-spline-1 performing worst and even reversing the direction of the association. For $\alpha_1$, modest bias and under-coverage were also observed, particularly for the SMRE and R-spline-1 approaches. Substantial bias arose in the estimates of $\gamma$ and $\sigma_e^2$ for these two approaches as well. As in the current curvature case, FPCA achieved the best performance in estimating $\sigma_e^2$, followed by P-spline. Figure S1 (top row) in the Supplementary Materials presents the LOOIC results for the different approaches under the two TB-WIV association settings. With the current curvature association, all approaches showed similar predictive performance, close to that of the true model. In contrast, under the cumulative curvature association, the less flexible SMRE and R-spline-1 approaches suffered a more pronounced loss in predictive accuracy compared with the other approaches.

\begin{table}[t]
\centering
\caption{Simulation results under Case~1 comparing DGP (true model), P-spline, FPCA, SMRE, and R-spline approaches (R-spline-1: one interior knot at midpoint; R-spline-2: three interior knots at equidistant quantiles) under current or cumulative curvature association. Parameters include the fixed effect in the survival submodel ($\gamma$), the association parameter for the current biomarker value ($\alpha_1$), the association for TB-WIV ($\alpha_2$), and the measurement error variance ($\sigma_e^2$). For each parameter, bias is computed from the posterior means across 100 data replications, and coverage probability (CP) is the proportion of times the 95\% credible interval contains the true parameter value across these replications.}
\label{tab:sim_case1}
\begingroup
\small
\setlength{\tabcolsep}{3pt}
\renewcommand{\arraystretch}{1.1}
\begin{tabular}{l*{6}{cc}}
\toprule
 & \multicolumn{2}{c}{DGP}
 & \multicolumn{2}{c}{P-spline}
 & \multicolumn{2}{c}{FPCA}
 & \multicolumn{2}{c}{SMRE}
 & \multicolumn{2}{c}{R-spline-1}
 & \multicolumn{2}{c}{R-spline-2} \\
\cmidrule(lr){2-3}\cmidrule(lr){4-5}\cmidrule(lr){6-7}\cmidrule(lr){8-9}\cmidrule(lr){10-11}\cmidrule(lr){12-13}
True value & Bias & CP
     & Bias & CP
     & Bias & CP
     & Bias & CP
     & Bias & CP
     & Bias & CP \\
\midrule
\multicolumn{1}{l}{\textbf{current curvature}}\\
$\gamma=-2$         & $-0.04$ & $93.0$ & $0.00$ & $99.0$ & $0.03$ & $98.0$ & $0.02$ & $97.0$ & $0.01$ & $97.0$ & $-0.03$ & $93.0$ \\
$\alpha_{1}=0.2$    & $0.01$  & $94.0$ & $0.00$ & $94.0$ & $0.00$ & $92.0$ & $0.05$ & $69.0$ & $0.04$ & $79.0$ & $0.00$  & $93.0$ \\
$\alpha_{2}=0.3$    & $0.00$  & $95.0$ & $-0.01$& $93.0$ & $-0.05$& $87.0$ & $-0.07$& $82.0$ & $-0.19$& $52.0$ & $-0.11$ & $13.0$ \\
$\sigma_e^2=1$      & $0.00$  & $91.0$ & $0.05$ & $10.0$ & $0.01$ & $89.0$ & $0.67$ & $0.0$  & $0.47$ & $0.0$  & $0.08$  & $1.0$  \\

\addlinespace[2pt]
\multicolumn{1}{l}{\textbf{cumulative curvature}}\\
$\gamma=-2$         & $-0.03$ & $93.0$ & $0.05$ & $92.0$ & $-0.01$ & $95.0$ & $0.27$ & $15.0$ & $0.28$ & $16.0$ & $-0.04$ & $94.0$ \\
$\alpha_{1}=0.2$    & $0.00$  & $96.0$ & $-0.03$& $86.0$ & $0.00$  & $93.0$ & $-0.02$& $79.0$ & $-0.03$& $82.0$ & $0.00$  & $94.0$ \\
$\alpha_{2}=0.3$    & $0.00$  & $91.0$ & $0.15$ & $41.0$ & $0.17$  & $11.0$ & $-0.25$& $0.0$  & $-0.59$& $0.0$  & $-0.13$ & $0.0$  \\
$\sigma_e^2=1$      & $0.00$  & $91.0$ & $0.05$ & $14.0$ & $0.01$  & $90.0$ & $0.66$ & $0.0$  & $0.46$ & $0.0$  & $0.07$  & $2.0$  \\
\bottomrule
\end{tabular}
\endgroup
\end{table}

\subsection{Case 2: model with non-spline-based trajectories}

This setting is adapted from the simulation study in \citet{wang2024modeling} to create a case where the true biomarker trajectory is no longer a spline function as in Case 1. More specifically, the subject-specific longitudinal trajectory combines the cubic drift function from Case 3 of \citet{wang2024modeling} with the sinusoidal component from their Case 4, and is given by
\begin{equation}
    \mu_i(t) \;=\; b_{i1} f_1(t) + b_{i2} + b_{i3} f_2(t), \qquad
    f_1(t) = -\tfrac{1}{6}(t-3)^3 + (t-3), \;\; f_2(t) = \sin(t),
\end{equation}
with $b_{i1} \sim \mathrm{Unif}(0.5,2)$, $b_{i2} \sim \mathrm{Unif}(4,8)$, and $b_{i3} \sim \mathrm{Unif}(0.5,1.5)$ drawn independently. Observed longitudinal data were generated over the time domain $[0,6]$, with visits scheduled every 0.4 time units from 0 to 2 and every 0.5 time units from 2.5 to 6. Measurement error variance was set to $\sigma_e^2=0.16$. No covariates were included in the longitudinal process.
The survival submodel follows the same structure as in our Case 1, with a single artificial covariate $w_{S,i}$ generated from a Bernoulli distribution with probability 0.5 of being 1 and associated coefficient $\gamma=-1$, and with association parameters $\alpha_1=\alpha_2=0.3$.
For the baseline hazard, as in our Case 1, we used a Weibull hazard with shape fixed at 3 for both TB-WIV variants. For the log-scale parameter, we set it to $-6.5$ under the current curvature formulation and to $-7.5$ under the cumulative curvature formulation, to achieve comparable event rates. Right-censoring times were generated from $\mathrm{Unif}(3,10)$, and follow-up was administratively censored at 6. Under these settings, the number of longitudinal measurements $n_i$ varied between 1 and 14, with a censoring rate of approximately 40\%. The sample size was fixed at $n=1000$.

Table \ref{tab:sim_case2} shows the estimation results, focusing on the same set of key model parameters as in Case 1. With the current curvature association, all approaches estimated $\gamma$ well and, except for the two R-spline-based approaches, $\alpha_1$ was also reasonably well estimated. For $\alpha_2$, P-spline, FPCA and SMRE achieved reasonable accuracy, whereas R-spline-2 performed worst and suffered significant attenuation. Estimation bias for $\sigma_e^2$ was generally small across approaches, despite poor coverage for SMRE and R-spline-2. Under the cumulative curvature association, all approaches showed satisfactory performance in estimating $\gamma$ and $\alpha_1$, but substantial bias arose for $\alpha_2$ across  approaches except for R-spline-1. For $\sigma_e^2$, the absolute bias was generally small, with  patterns similar to those observed under the current curvature association. 
Figure S1 (middle row) in the Supplementary Materials presents the LOOIC results under the two TB-WIV association settings. Overall, the approaches delivered broadly comparable predictive performance, close to that of the true model. The only clear difference was under the cumulative curvature association, where R-spline-2 tended to show comparatively lower predictive accuracy than the other approaches.

\begin{table}[t]
\centering
\caption{Simulation results under Case 2 comparing DGP (true model), P-spline, FPCA, SMRE, and R-spline approaches (R-spline-1: one interior knot at midpoint; R-spline-2: three interior knots at equidistant quantiles) under current or cumulative curvature association. Parameters include the fixed effect in the survival submodel ($\gamma$), the association parameter for the current biomarker value ($\alpha_1$), the association for TB-WIV ($\alpha_2$), and the measurement error variance ($\sigma_e^2$). Other settings are the same as in Table \ref{tab:sim_case1}.}
\label{tab:sim_case2}
\begingroup
\small
\setlength{\tabcolsep}{3pt}
\renewcommand{\arraystretch}{1.1}
\begin{tabular}{l*{6}{cc}}
\toprule
 & \multicolumn{2}{c}{DGP}
 & \multicolumn{2}{c}{P-spline}
 & \multicolumn{2}{c}{FPCA}
 & \multicolumn{2}{c}{SMRE}
 & \multicolumn{2}{c}{R-spline-1}
 & \multicolumn{2}{c}{R-spline-2} \\
\cmidrule(lr){2-3}\cmidrule(lr){4-5}\cmidrule(lr){6-7}\cmidrule(lr){8-9}\cmidrule(lr){10-11}\cmidrule(lr){12-13}
True value & Bias & CP
     & Bias & CP
     & Bias & CP
     & Bias & CP
     & Bias & CP
     & Bias & CP \\
\midrule
\multicolumn{1}{l}{\textbf{current curvature}}\\
$\gamma=-1$         & $0.00$ & $98.0$ & $0.00$ & $97.0$ & $0.01$ & $98.0$ & $0.00$ & $97.0$ & $0.01$ & $97.0$ & $0.02$ & $97.0$ \\
$\alpha_{1}=0.3$    & $0.00$ & $97.0$ & $-0.01$& $94.0$ & $-0.01$& $96.0$ & $0.00$ & $97.0$ & $-0.06$& $60.0$ & $-0.05$& $68.0$ \\
$\alpha_{2}=0.3$    & $0.00$ & $97.0$ & $0.03$ & $91.0$ & $0.00$ & $95.0$ & $-0.03$& $86.0$ & $0.05$ & $89.0$ & $-0.09$& $3.0$  \\
$\sigma_e^2=0.16$   & $0.00$ & $95.0$ & $0.00$ & $96.0$ & $0.00$ & $64.0$ & $0.01$ & $48.0$ & $0.00$ & $97.0$ & $-0.01$& $8.0$  \\

\addlinespace[2pt]
\multicolumn{1}{l}{\textbf{cumulative curvature}}\\
$\gamma=-1$         & $-0.01$ & $96.0$ & $-0.04$& $92.0$ & $-0.02$& $97.0$ & $-0.01$& $97.0$ & $-0.01$& $97.0$ & $-0.01$& $96.0$ \\
$\alpha_{1}=0.3$    & $0.00$  & $95.0$ & $0.01$ & $94.0$ & $0.01$ & $95.0$ & $0.00$ & $95.0$ & $0.00$ & $97.0$ & $-0.01$& $97.0$ \\
$\alpha_{2}=0.3$    & $0.00$  & $98.0$ & $0.23$ & $18.0$ & $0.19$ & $11.0$ & $-0.08$& $24.0$ & $0.00$ & $100.0$& $-0.20$& $1.0$  \\
$\sigma_e^2=0.16$   & $0.00$  & $94.0$ & $0.00$ & $96.0$ & $0.00$ & $61.0$ & $0.01$ & $29.0$ & $0.00$ & $96.0$ & $-0.01$& $13.0$ \\
\bottomrule
\end{tabular}
\endgroup
\end{table}

\subsection{Case 3: model motivated by UK CF data}

This setting aims to mimic the empirical patterns observed in our application data. To this end, for each TB-WIV definition we fitted the joint model specified in our real data analysis (Section \ref{sec:app}), with the subject-specific trajectory function $\mu_i(t)$ represented using the R-spline-based approach. To determine the knot setting, we considered three configurations with $1/2/3$ interior knots placed at equidistant quantiles of the observed follow-up times, and the final model was selected based on the LOOIC. For both TB-WIV definitions, the model with a single interior knot was selected, and we used the fitted parameters from this model as the basis for simulation. 
The observation design also resembled the real data pattern. Delayed entry times $T_{\text{start},i}$ were drawn from a truncated log-normal distribution on $(0,31.9)$, with parameters obtained by fitting the distribution to the real data. To generate the three baseline covariates considered in the application, two binary covariates (diagnosis and genotype indicators) were generated from Bernoulli distributions with success probabilities $0.45$ and $0.483$, respectively, and a third covariate defined as age at entry since 18 (per decade), i.e., $T_{\text{start},i}/10$. Follow-up times were generated irregularly with gaps between $0.9$ and $1.6$ years, and right-censoring times were given by $T_{\text{start},i}+\mathrm{Unif}(3,28)$, with administrative censoring at $32$.  Under these settings, the number of measurements per subject typically varied between 1 and 24, with a median around 9, and the censoring rate was approximately $70\%$, both close to those observed in the real data.

Table \ref{tab:sim_case3} shows the estimation results for fixed-effect parameters, association parameters, and the measurement error variance. Under the current curvature association, R-spline-1 achieved good performance except for poor coverage of $\sigma_e^2$. FPCA and P-spline gave satisfactory performance across most parameters, but both showed attenuation for $\alpha_2$ (with a stronger effect observed for P-spline), which was the most challenging parameter to estimate across approaches. SMRE showed moderate bias for both $\alpha_1$ and $\alpha_2$ and the worst performance for $\sigma_e^2$. R-spline-2 yielded the largest bias for $\alpha_2$ and moderate bias for some parameters in the longitudinal submodel.
Under the cumulative curvature association, P-spline and FPCA achieved satisfactory performance for most parameters except for $\alpha_2$, with FPCA suffering a more pronounced positive bias and loss of coverage. 
No approach could accurately estimate $\alpha_2$ in this case. SMRE demonstrated moderate bias for both $\alpha_1$ and $\alpha_2$, and the poorest performance for $\sigma_e^2$ among the approaches. R-spline-1 performed well except for $\alpha_2$ and $\sigma_e^2$, while R-spline-2 suffered from varying levels of bias across most parameters.
Table S1 in the Supplementary Materials shows the results obtained when we set the TB-WIV (current or cumulative curvature) association to zero at the data-generating stage, with all other settings unchanged. For both TB-WIV definitions, all approaches, except for R-spline-2, detected the null effect relatively accurately, in the sense that coverage probabilities for $\alpha_2=0$ were close to the $95\%$ nominal level. Estimation of the other parameters followed similar patterns as in the non-null case. With regard to predictive performance, as shown in Figure S1 (bottom row) of the Supplementary Materials, no approach exhibited a noticeable loss in LOOIC relative to the true model under either type of TB-WIV definition.

\begin{table}[t]
\centering
\caption{Simulation results under Case~3 comparing DGP (true model), P-spline, FPCA, SMRE, and R-spline approaches (R-spline-1: one interior knot at midpoint; R-spline-2:  three interior knots at equidistant quantiles) under current or cumulative curvature association. Parameters include the fixed effects in the longitudinal submodel ($\beta_1, \beta_2, \beta_3$), those in the survival submodel ($\gamma_1, \gamma_2, \gamma_3$), the association parameter for the current biomarker value ($\alpha_1$), the association for TB-WIV ($\alpha_2$), and the measurement error variance ($\sigma_e^2$). Other settings are the same as in Table \ref{tab:sim_case1}.}
\label{tab:sim_case3}
\begingroup
\small
\setlength{\tabcolsep}{3pt}
\renewcommand{\arraystretch}{1.1}
\begin{tabular}{l*{6}{cc}}
\toprule
 & \multicolumn{2}{c}{DGP}
 & \multicolumn{2}{c}{P-spline}
 & \multicolumn{2}{c}{FPCA}
 & \multicolumn{2}{c}{SMRE}
 & \multicolumn{2}{c}{R-spline-1}
 & \multicolumn{2}{c}{R-spline-2} \\
\cmidrule(lr){2-3}\cmidrule(lr){4-5}\cmidrule(lr){6-7}\cmidrule(lr){8-9}\cmidrule(lr){10-11}\cmidrule(lr){12-13}
True value & Bias & CP
     & Bias & CP
     & Bias & CP
     & Bias & CP
     & Bias & CP
     & Bias & CP \\
\midrule
\multicolumn{1}{l}{\textbf{current curvature}}\\
$\beta_{1}=0.25$    & $0.00$  & $89.0$ & $0.00$  & $95.0$ & $-0.01$ & $90.0$ & $0.00$  & $94.0$ & $0.00$  & $92.0$ & $-0.01$ & $80.0$ \\
$\beta_{2}=-0.19$   & $0.00$  & $96.0$ & $0.01$  & $92.0$ & $0.01$  & $93.0$ & $0.01$  & $93.0$ & $0.00$  & $96.0$ & $0.01$  & $81.0$ \\
$\beta_{3}=0.13$    & $0.00$  & $92.0$ & $0.04$  & $86.0$ & $-0.01$ & $86.0$ & $-0.05$ & $86.0$ & $0.00$  & $91.0$ & $-0.04$ & $69.0$ \\
$\gamma_{1}=-0.19$  & $0.00$  & $95.0$ & $0.00$  & $96.0$ & $0.00$  & $95.0$ & $0.05$  & $94.0$ & $0.00$  & $97.0$ & $0.00$  & $97.0$ \\
$\gamma_{2}=0.02$   & $-0.02$ & $95.0$ & $-0.03$ & $95.0$ & $-0.02$ & $94.0$ & $-0.06$ & $92.0$ & $-0.02$ & $94.0$ & $-0.01$ & $97.0$ \\
$\gamma_{3}=0.04$   & $0.01$  & $96.0$ & $0.04$  & $97.0$ & $0.03$  & $96.0$ & $0.08$  & $95.0$ & $0.03$  & $96.0$ & $0.00$  & $96.0$ \\
$\alpha_{1}=-2.95$  & $-0.04$ & $97.0$ & $-0.10$ & $92.0$ & $0.06$  & $94.0$ & $-0.26$ & $61.0$ & $0.00$  & $96.0$ & $0.15$  & $81.0$ \\
$\alpha_{2}=21.46$  & $-2.08$ & $91.0$ & $-7.70$ & $31.0$ & $-4.65$ & $68.0$ & $-4.76$ & $80.0$ & $1.33$  & $99.0$ & $-10.48$& $0.0$  \\
$\sigma_e^2=0.04$   & $0.00$  & $94.0$ & $0.00$  & $97.0$ & $0.00$  & $91.0$ & $0.01$  & $0.0$  & $0.00$  & $3.0$  & $0.00$  & $57.0$ \\

\addlinespace[2pt]
\multicolumn{1}{l}{\textbf{cumulative curvature}}\\
$\beta_{1}=0.24$    & $0.00$  & $95.0$ & $0.00$  & $93.0$ & $-0.01$ & $93.0$ & $0.00$  & $91.0$ & $0.00$  & $96.0$ & $-0.01$ & $85.0$ \\
$\beta_{2}=-0.20$   & $0.00$  & $97.0$ & $0.01$  & $94.0$ & $0.01$  & $95.0$ & $0.01$  & $93.0$ & $0.01$  & $96.0$ & $0.02$  & $83.0$ \\
$\beta_{3}=0.13$    & $0.00$  & $93.0$ & $0.04$  & $92.0$ & $-0.01$ & $87.0$ & $-0.05$ & $91.0$ & $0.00$  & $91.0$ & $-0.04$ & $63.0$ \\
$\gamma_{1}=-0.19$  & $-0.02$ & $96.0$ & $-0.02$ & $96.0$ & $-0.04$ & $96.0$ & $0.01$  & $94.0$ & $-0.02$ & $96.0$ & $-0.03$ & $99.0$ \\
$\gamma_{2}=0.01$   & $-0.01$ & $93.0$ & $-0.02$ & $93.0$ & $0.01$  & $96.0$ & $-0.04$ & $92.0$ & $0.00$  & $95.0$ & $-0.02$ & $95.0$ \\
$\gamma_{3}=-0.17$  & $-0.02$ & $93.0$ & $0.05$  & $95.0$ & $-0.04$ & $92.0$ & $0.13$  & $87.0$ & $0.06$  & $93.0$ & $-0.48$ & $37.0$ \\
$\alpha_{1}=-3.06$  & $-0.05$ & $95.0$ & $-0.09$ & $95.0$ & $0.07$  & $88.0$ & $-0.19$ & $79.0$ & $-0.01$ & $94.0$ & $-0.26$ & $73.0$ \\
$\alpha_{2}=4.85$   & $0.05$  & $97.0$ & $1.41$  & $84.0$ & $3.71$  & $24.0$ & $-0.90$ & $82.0$ & $3.68$  & $26.0$ & $-0.75$ & $81.0$ \\
$\sigma_e^2=0.04$   & $0.00$  & $93.0$ & $0.00$  & $93.0$ & $0.00$  & $93.0$ & $0.01$  & $0.0$  & $0.00$  & $3.0$  & $0.00$  & $63.0$ \\
\bottomrule
\end{tabular}
\endgroup
\end{table}

\subsection{Summary and discussion of simulation results}  
\label{subsec:sim_disc}

Across the three cases, focusing on the overall inference performance on the model parameters, P-spline and FPCA both generally performed best across parameters, with results close to the true model. SMRE performed somewhat less well overall in comparison to P-spline and FPCA, and the R-spline approach, especially the three-knot variant, was the least reliable. Inference on the association parameters, especially the TB-WIV association ($\alpha_2$), was more challenging than for other parameters. Under the current curvature association, FPCA and P-spline generally provided the best or near-best estimation performance compared with the alternative approaches. Under cumulative curvature, estimation became difficult for all approaches: P-spline and FPCA tended to show positive bias, SMRE typically showed attenuation, whereas the R-spline approaches showed more variable patterns depending on the knot setting. When the TB-WIV association (current or cumulative curvature) was absent, the null effect was more reliably detected by all approaches except R-spline (the three-knot version). For the measurement error variance, FPCA and P-spline were the only approaches that produced accurate or near-accurate inference across settings, a precondition for cleanly separating measurement error from within-trajectory biological variation. Although P-spline showed under-coverage in Case 1 and FPCA in Case 2, the associated bias was very small and unlikely to be of practical concern. SMRE consistently performed poorest, which may be unsurprising given its limited flexibility in modelling the trajectory patterns. Compared with other parameters, the fixed-effect coefficients $(\beta,\gamma)$ were inferred more robustly by all approaches, with P-spline and FPCA generally providing the best accuracy.

To understand why the TB–WIV association is challenging to infer in general settings, even with flexible P-spline or FPCA-based approaches, note that the same underlying basis used to represent the mean trajectory also determines its curvature (via differentiation of that basis), and estimation is dominated by the longitudinal likelihood, which favours components that best reconstruct subject-specific trajectories. Consequently, curvature-oriented directions can be weakly identified under sparse longitudinal sampling.
For FPCA, its basis functions (i.e., eigenfunctions) are fully data-driven. As explained in Section \ref{subsec:methods-WIV-fpca}, FPCA prioritises the directions that explain the largest share of total variability in the trajectories. Therefore we expect FPCA to perform better when the curvature pattern is contained in the leading, high-variance components, and worse when it sits in low-variance components that are subject to noise and shrinkage.
For the P-spline approach, although the basis functions are not data-adaptive as in the FPCA, they are ordered by roughness by construction: those capturing more local, wiggly behaviour have smaller $L^2$-amplitude (due to stronger penalties). Therefore curvature-related components tend to be weakly identified, and their associated coefficients are more prone to shrinkage during estimation.

The two specific TB-WIV definitions considered in this paper have different implications for estimation performance owing to their distinct characteristics. For the current curvature, $ \lvert \mu_i''(t) \rvert $, the quantity is pointwise, depending on the second derivative at time $t$. It is therefore less susceptible to sustained deviations or systematic bias in $ \lvert \mu_i''(t) \rvert $ over any sub-interval, but more susceptible to large random pointwise errors, especially near the event time. This may partly explain why the current curvature association is more prone to attenuation: pointwise estimation of curvature can be noisy, although the attenuation is often mild for P-spline and FPCA owing to their generic flexibility.
For the cumulative curvature, note that $\sqrt{\int_{0}^{t} (\mu_i''(s))^{2}\,ds} \approx \sqrt{\Delta_t \sum_{m=1}^{M} (\mu_i''(t_m))^{2}}$  by a Riemann-sum approximation. Thus it depends on the aggregation of $(\mu_i''(t_m))^{2}$ over $[0,t]$, and is therefore more susceptible to sustained deviations or systematic bias in $\lvert \mu_i''(t) \rvert$ over any sub-interval,  and less susceptible to random pointwise errors due to the possible cancellation effect over time. In our simulations, a positive bias in the estimated association for the cumulative measure was often seen with FPCA and P-spline. While both approaches can, in general, approximate the population-level curvature well, weak identification at the subject level induces shrinkage of individual cumulative curvature measures towards the population mean. As a result, subjects with larger cumulative curvature tend to be underestimated, and vice versa. Because the measure is cumulative, this shrinkage bias can get locked in and be carried forward to later $t$, creating systematic biases. This phenomenon can be seen more clearly through a simple working model. At a given $t$, suppose that the estimated and true cumulative curvature can be related by a shrinkage model 
\begin{equation}\label{eq:shrinkage-cumcurv}
\widehat{\mathrm{WIV}}_i(t) = a(t) + k(t)\,\mathrm{WIV}_i(t) + \epsilon_i(t).
\end{equation}
If $0<k(t)\approx k<1$ (representing a time-independent shrinkage effect) and $\epsilon_i(t)$ is negligible, then this translates to an inflated association of $\alpha_2$ by a factor $1/k$.
Note that in the presence of left truncation, we expect differences between FPCA and P-spline in estimating the current or cumulative curvature association due to their distinct basis characteristics and extrapolation behaviour (see Figure \ref{fig:app_long} for an illustration). The P-spline approach tends to impose similar wiggly behaviour across the domain due to the quasi-periodic nature of the underlying basis. This may increase pointwise ‘noise’ in the estimated curvature over the extrapolated region and the scale of the cumulative curvature, which could result in additional attenuation of the estimated association (compared to FPCA). For cumulative curvature, such attenuation may partly offset the usual positive bias (as seen in Case 3), though it may overshoot; the precise behaviour is unfortunately unknown a priori and setting-dependent.

With respect to predictive performance, as measured by survival-based LOOIC, differences among approaches were less prominent than for inference under both TB-WIV definitions, and FPCA and P-spline-based approaches generally yielded results close to the true model across most settings. This pattern may be partly explained by a possible compensation effect within the survival submodel, and a more precise evaluation of this impact depends on the error structure of the estimated TB-WIV. For instance, under the working model in \eqref{eq:shrinkage-cumcurv}, together with our hazard specification in \eqref{eq:jm-wiv-surv}, predictive power may be largely retained if $k(t)$ is approximately constant over $t$, the baseline hazard is sufficiently flexible to absorb $a(t)$, and $\epsilon_i(t)$ is either small or time-constant. Violations of these conditions, depending on the degree of departure, may result in a loss of predictive performance. Another factor influencing the degree of such loss relates to the intrinsic strength of the TB-WIV signal in the true model, with the impact expected to be larger when the magnitude of the TB-WIV association ($|\alpha_2|$) and the between-subject variability of the TB-WIV feature are greater.

\section{Application to the UK CF data}
\label{sec:app}
\begin{figure}[t]
\centering
\includegraphics[width=0.48\linewidth]{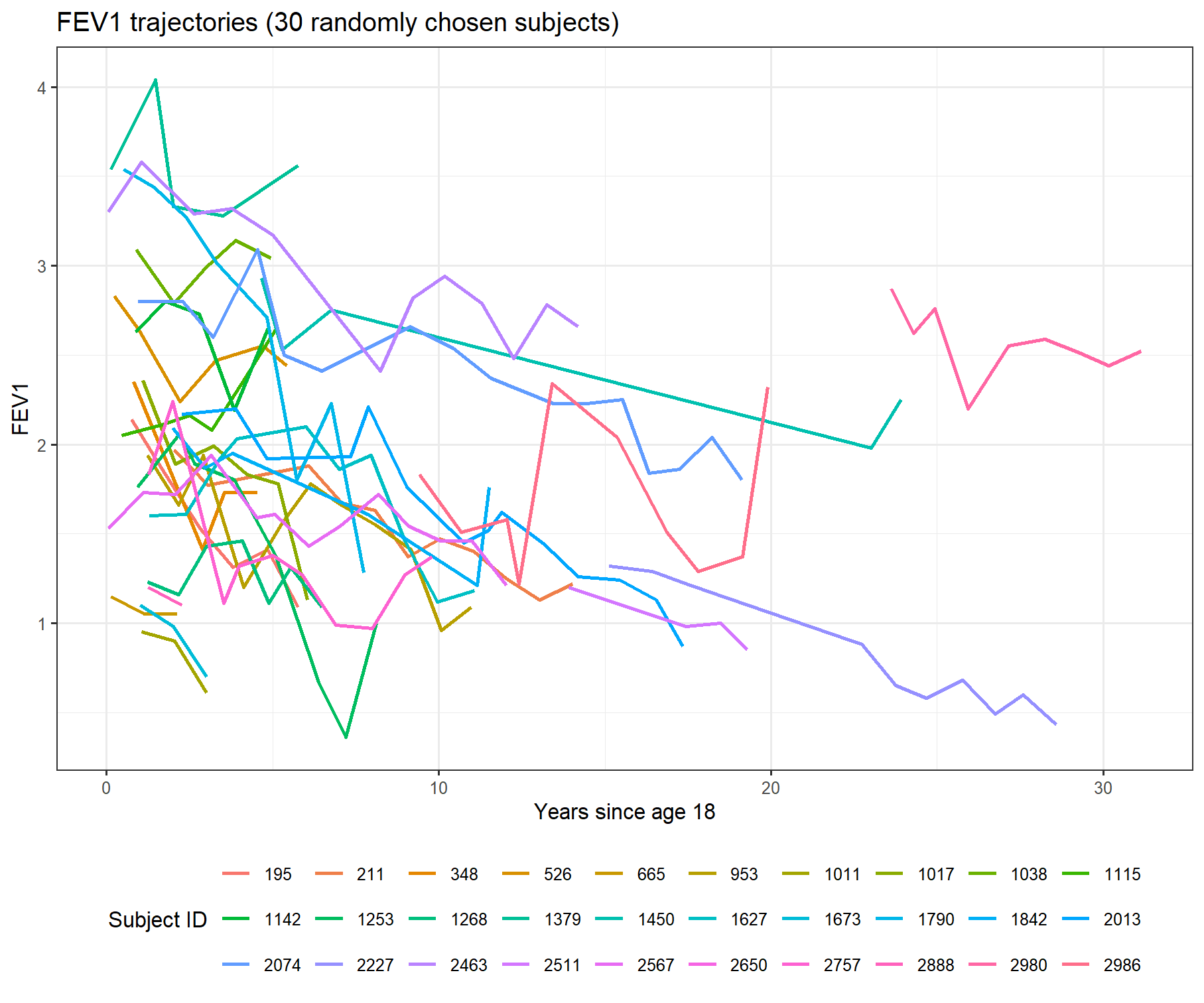}
\includegraphics[width=0.48\linewidth]{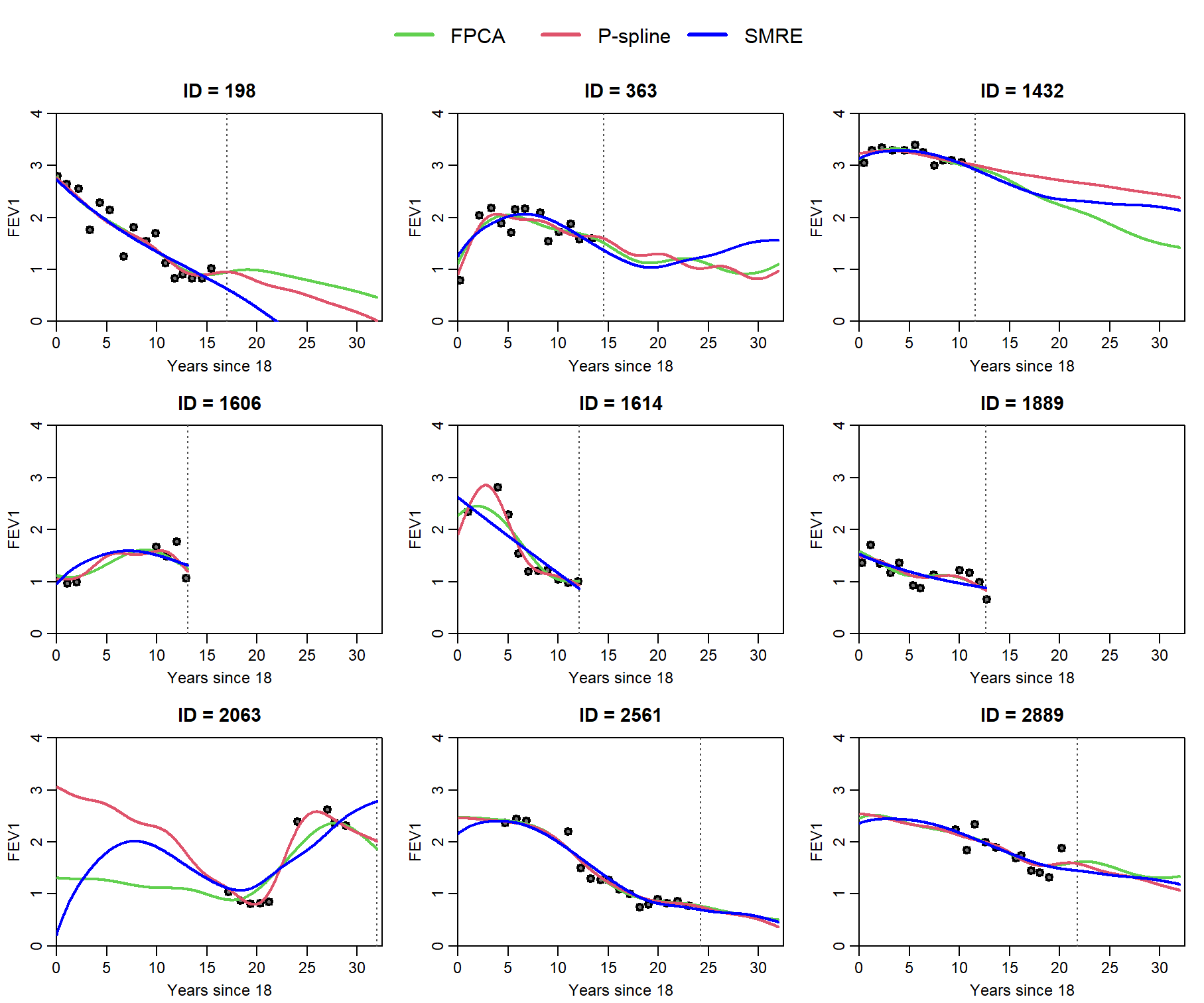}
\caption{Left panel: observed FEV1 trajectories for 30 randomly selected subjects from the UK CF dataset. Right panel: fitted trajectories for nine randomly selected subjects under the three approaches: FPCA (green), P-spline (red), and SMRE (blue), conditional on posterior mean estimates and the cumulative curvature association. Black dots denote observed FEV1 measurements, and solid lines show the fitted trajectories under each approach; for censored subjects, trajectories are extrapolated beyond the last observation. The vertical dotted line indicates the observed follow-up time.}
\label{fig:app_long}
\end{figure}

Our analysis dataset is drawn from the UK Cystic Fibrosis Registry, a national, anonymised, longitudinal database that has systematically collected annual clinical data on people with CF since 1996. The registry covers more than $99\%$ of the UK CF population and includes annual measurements of lung function, anthropometrics, and major clinical outcomes.
For our analysis, we considered a subcohort of adult women aged 18--49 years, observed between 1996 and 2020, who had not received a lung transplant and had complete baseline data for genotype and age at diagnosis (detailed definitions given below). These inclusion criteria follow \citet{palma2025bayesian} and \citet{palma2024demographic}, and we refer to their papers for further details. The resulting cohort comprised $N=3282$ women with a total of $29094$ annual encounters. We did not include men, as the women-only subgroup already provides a sufficiently large sample for analysis and is clinically motivated as survival patterns differ by sex in the UK CF population, with women consistently showing lower life expectancy than men \citep{palma2024demographic}.
As in \citet{palma2025bayesian}, follow-up was defined as the time in years from age 18 to either death (if no transplant) or censoring at transplant, age 50 or study end (whichever comes first), resulting in a censoring rate of $73.9\%$. The biomarker of interest is FEV1, a key lung function indicator (in litres), with higher values indicating better lung function. The number of measurements per individual ranged from 1 to 25 during follow-up, with a median of 8 and an interquartile range of $[5,12]$. Figure \ref{fig:app_long} (left panel) displays longitudinal trajectories of FEV1 for a random sample of 30 patients, illustrating the substantial heterogeneity in trajectory shapes and measurement schedules, with subjects entering follow-up at different ages.

Our interest lies in assessing the association between TB-WIV of FEV1 and the risk of death in individuals with CF, using the joint modelling framework described in Section~\ref{subsec:JM-WIV-formulation}. 
We considered three TB-WIV definitions: current curvature; cumulative curvature; and cumulative curvature with a one-year window, that is, $\text{WIV}_i(t)=\sqrt{\int_{\max(0,\,t-1)}^{t}\{\mu_i''(s)\}^2\,ds}$. The latter windowed version is a special case of the weighted cumulative curvature measure with a two-piece step (piecewise-constant) weighting function, $\omega_\sigma(t-s)=\mathbf{1}\{\,s\in[\max(0,\,t-1),\,t]\,\}$.
As in \citet{palma2025bayesian}, the time scale in both submodels was set to chronological age since 18 years to naturally capture age-related effects.
We included three baseline covariates in both submodels ($W_{L,i}$ and $W_{S,i}$). The first two followed \citet{palma2025bayesian}: indicators for diagnosis after the first year of life (reference: $\leq 1$ year) and homozygous F508del genotype (reference: non-homozygous). The diagnosis indicator separates infant from later diagnosis and is commonly considered in CF studies due to the potential implications for differences in phenotype and disease severity. F508del homozygosity is the most common CF-causing CFTR mutation and is known to be prognostic. We also included age at entry since 18 (per decade) as an additional covariate to adjust for potential entry-time effects.
To model the subject-specific trajectory function $\mu_i(t)$, we considered the three semiparametric approaches described in Section \ref{sec:methods-WIV} and compared their performance. For the survival submodel, we flexibly modelled the logarithm of the baseline hazard function using cubic B-splines with a single interior knot placed at the median of the observed event times.
Posterior inference was based on the same set of weakly informative priors as used in the simulation study, with the likelihood function adjusted in the survival part to account for left truncation. Conditional on the TB-WIV definition and the chosen modelling approach for $\mu_i(t)$, MCMC sampling for the resulting joint model was implemented in \texttt{RStan} as in the simulation study, using two independent chains, each with 1000 iterations for burn-in and a further 1000 retained for inference. Convergence diagnostics indicated satisfactory performance across all settings.
The total computational time, including burn-in, for each combination of approach and TB-WIV definition is reported in Table S2 of the Supplementary Materials.

We examined fitted trajectories from the three approaches across individuals (see right panel of Figure \ref{fig:app_long} for randomly selected examples under cumulative curvature; results under the alternative definition were almost identical). The patterns are consistent with expectations and with our simulation findings (Case 3). Both FPCA- and P-spline-based approaches showed greater flexibility in capturing individual trajectories than the SMRE-based approach. Relative to FPCA, P-splines tended to produce more wiggly fits and extrapolations, while SMRE gave flatter trajectories, particularly when data were sparse. We also empirically assessed the sufficient working condition for SMRE, as described in \citet{luo2025semiparametric}, and observed some deviation. 

Table \ref{tab:realdata_est} shows estimation results obtained from the P-spline and FPCA-based approaches under the three TB-WIV definitions. SMRE yielded broadly similar overall patterns, with results provided in Table S3 of the Supplementary Materials and not discussed further here. For the longitudinal parameters, estimates were nearly identical across TB-WIV specifications, and the two modelling approaches also yielded relatively similar results within each specification. As expected, and consistent with results reported in \citet{palma2025bayesian}, later diagnosis (vs.\ $\leq$1 year) was associated with higher FEV1 (better lung function), whereas F508del homozygosity (vs.\ non-homozygous) was associated with lower FEV1 (worse lung function). Entry age was also found to be a significant predictor of FEV1 (95\% CI excluded 0), with later entry associated with higher FEV1. This is reasonable to expect as it likely reflects healthy-survivor selection effect. For measurement error variance, P-spline yielded a slightly lower estimate than FPCA, which is consistent with the patterns seen in the fitted trajectories.

The two modelling approaches also gave similar inferences on parameters in the survival submodel under the three TB-WIV definitions, except for $\alpha_2$. For the baseline covariates, after adjusting for the underlying FEV1 level and TB-WIV, F508del homozygosity and entry age show no additional direct effect on the hazard (95\% CIs for $\gamma_2$ and $\gamma_3$ include 0). By contrast, diagnosis after age 1 retains a significant direct association, with the negative sign of $\gamma_1$ indicating a lower hazard for those diagnosed later, which is biologically reasonable. The association parameter for the underlying FEV1 level ($\alpha_1$) shows a significant negative association, which aligns with results in \citet{palma2025bayesian}, indicating that worse lung function is linked to higher risk of death.
TB-WIV, whether defined through current or cumulative curvature, showed highly significant positive associations with the risk of death under both approaches ($\alpha_2$), indicating that higher TB-WIV is linked to increased mortality. This finding is novel, yet consistent with existing literature and clinical expectations. The current curvature and the cumulative curvature with a one-year window yielded very similar association estimates, which is expected given the approximate relationship between these measures noted earlier.
Conditional on the TB-WIV definition, it is noteworthy that the P-spline and FPCA approaches yielded $\alpha_2$ estimates of very different magnitudes, with P-splines producing smaller effects, particularly under cumulative curvature. This pattern, which was also observed under Case 3 in the simulation study, likely results from attenuation due to the extra noise introduced by the slightly more wiggly P-spline fits. The attenuation effect here is likely to be stronger than in the simulations, due to additional noise and irregularities in the real data.

Table \ref{tab:realdata_loo} compares the predictive performance for the survival outcome under different JM settings and modelling approaches using LOOIC. For all approaches, incorporating TB-WIV improves predictive performance. When TB-WIV is included in the joint model, both FPCA- and P-spline-based approaches yield superior performance compared with SMRE, with FPCA under cumulative curvature achieving the best performance among the different configurations. The P-spline approach appears to suffer some loss of predictive power under cumulative curvature, likely due to the attenuation effect noted earlier.
As a further comparison, we also fitted the JM with RB-WIV following \citet{palma2025bayesian}, using the same set of baseline covariates and baseline hazard specification as in our model (see Section S1 of the Supplementary Materials for details). In contrast to \citet{palma2025bayesian}, we included a random slope for modelling the subject-specific trajectory (omitted in their work due to computational complexity) to allow a fairer comparison with our approaches. The LOOIC estimated with the resulting RB-WIV model was 6115.9, which is inferior to that obtained under the cumulative curvature specification with FPCA. The corresponding z-score, obtained by dividing the estimated difference in LOOIC by its standard error, yielded a magnitude of 2.48, providing non-trivial evidence in favour of the cumulative curvature specification. These results suggest that, for the present CF dataset with FEV1 as the biomarker, joint models with cumulative curvature association may offer better survival predictive performance than their RB-WIV counterparts. A more detailed and systematic comparison, however, is beyond the scope of this paper.

\begin{sidewaystable}[!htp]
\centering
\caption{Estimation results obtained from the P-spline and FPCA-based approaches under three TB-WIV specifications: current curvature; cumulative curvature; and cumulative curvature with a one-year window. Parameters include fixed effects ($\beta_1, \beta_2, \beta_3$ and $\gamma_1, \gamma_2, \gamma_3$), measurement error variance ($\sigma_e^2$), and association parameters ($\alpha_1, \alpha_2$). Posterior means are reported as point estimates (the corresponding posterior distributions are approximately symmetric). The 95\% credible intervals (CIs), shown in brackets, are based on the 2.5th and 97.5th percentiles of the posterior samples.}
\label{tab:realdata_est}
\begingroup
\small
\setlength{\tabcolsep}{2.5pt}
\renewcommand{\arraystretch}{1.1}
\begin{tabular}{lcccccc}
\toprule
 & \multicolumn{2}{c}{Current curvature} & \multicolumn{2}{c}{Cumulative curvature} & \multicolumn{2}{c}{Cumulative curvature (one-year window)} \\
\cmidrule(lr){2-3}\cmidrule(lr){4-5}\cmidrule(lr){6-7}
Parameter & P-spline & FPCA & P-spline & FPCA & P-spline & FPCA \\
\midrule
\multicolumn{1}{l}{\textbf{Longitudinal}} \\
$\beta_{1}$ -- Diagnosed after year 1     & 0.25 (0.18, 0.30)    & 0.26 (0.20, 0.32)    & 0.24 (0.19, 0.30)    & 0.27 (0.22, 0.33)    & 0.25 (0.19, 0.31)    & 0.27 (0.22, 0.32) \\
$\beta_{2}$ -- F508del homozygous         & -0.17 (-0.23, -0.12) & -0.23 (-0.29, -0.18) & -0.18 (-0.23, -0.12) & -0.23 (-0.28, -0.17) & -0.17 (-0.23, -0.11) & -0.23 (-0.28, -0.18) \\
$\beta_{3}$ -- Entry age (10y)            & 0.21 (0.16, 0.26)    & 0.16 (0.12, 0.20)    & 0.21 (0.16, 0.26)    & 0.16 (0.12, 0.20)    & 0.21 (0.16, 0.26)    & 0.16 (0.12, 0.20) \\
$\sigma_e^2$ -- Measurement error variance         & 0.03 (0.03, 0.03)    & 0.04 (0.04, 0.04)    & 0.03 (0.03, 0.03)    & 0.04 (0.04, 0.04)    & 0.03 (0.03, 0.03)    & 0.04 (0.04, 0.04) \\
\addlinespace[2pt]
\multicolumn{1}{l}{\textbf{Survival}} \\
$\gamma_{1}$ -- Diagnosed after year 1    & -0.17 (-0.32, -0.02) & -0.16 (-0.32, 0.00)  & -0.19 (-0.34, -0.03) & -0.18 (-0.38, 0.01)  & -0.17 (-0.33, -0.01) & -0.17 (-0.32, -0.02) \\
$\gamma_{2}$ -- F508del homozygous        & 0.03 (-0.12, 0.17)   & 0.01 (-0.14, 0.16)   & 0.03 (-0.12, 0.19)   & 0.01 (-0.15, 0.18)   & 0.03 (-0.12, 0.17)   & 0.01 (-0.14, 0.15) \\
$\gamma_{3}$ -- Entry age (10y)           & 0.13 (-0.06, 0.32)   & 0.09 (-0.10, 0.28)   & 0.14 (-0.05, 0.34)   & -0.15 (-0.39, 0.08)  & 0.13 (-0.06, 0.32)   & 0.09 (-0.10, 0.28) \\
$\alpha_{1}$ -- Association (mean)        & -3.19 (-3.41, -2.97) & -3.10 (-3.33, -2.90) & -3.18 (-3.40, -2.96) & -3.63 (-4.01, -3.30) & -3.19 (-3.41, -2.97) & -3.11 (-3.33, -2.89) \\
$\alpha_{2}$ -- Association (TB-WIV)      & 4.98 (3.59, 6.39)    & 17.81 (13.49, 21.83) & 1.52 (0.79, 2.31)    & 11.34 (8.52, 14.44)  & 5.29 (3.81, 6.77)    & 17.31 (12.89, 21.90) \\
\bottomrule
\end{tabular}
\endgroup
\end{sidewaystable}

\begin{table}[t]
\centering
\caption{LOOIC for the survival outcome under different joint model specifications and modelling approaches. Lower values indicate better predictive performance.}
\label{tab:realdata_loo}
\begingroup
\small
\setlength{\tabcolsep}{6pt}
\renewcommand{\arraystretch}{1.1}
\begin{tabular}{lccc}
\toprule
Joint model setting & FPCA & P-spline & SMRE \\
\midrule
no TB-WIV            & 6192.9 & 6134.3 & 6182.1 \\
current curvature    & 6122.3 & 6078.0 & 6152.3 \\
cumulative curvature (one-year window) & 6128.5 & 6079.5 &  6151.1 \\
cumulative curvature & 6046.1 & 6102.5 & 6132.0 \\
\bottomrule
\end{tabular}
\endgroup
\end{table}

\section{Discussion}
\label{sec:disc}

In this paper, we explored three Bayesian semiparametric approaches, namely P-spline, FPCA and SMRE-based methods, for modelling longitudinal biomarker trajectories and TB-WIV in the joint modelling framework. The P-spline and FPCA approaches are introduced here for the first time in this study context, while the SMRE approach is extended from the recent proposal in \citet{luo2025semiparametric}. Motivated by TB-WIV measures proposed in the literature, we focused on two specific definitions of TB-WIV, current curvature and cumulative curvature. In terms of estimating the TB-WIV association, with current curvature, P-spline and FPCA tend to yield inference closer to the true model, whereas with cumulative curvature different bias patterns emerge across approaches. FPCA and P-spline often yield similar estimates, although under delayed entry the P-spline approach may suffer attenuation. We also examined the performance of the three approaches in detecting a null effect of both versions of TB-WIV, and we found it can be detected reasonably reliably by all three approaches. Overall, the more flexible FPCA and P-spline approaches provide more accurate estimates of most model parameters than SMRE. The R-spline approach, included as a benchmark, tends to be the least reliable and risks serious estimation bias and coverage problems, with performance strongly dependent on the knot setting. For survival prediction, as measured by LOOIC, differences between approaches are less pronounced, and loss in performance is often minimal. We investigated the association between three definitions of TB-WIV (current curvature, cumulative curvature with a one-year window, and cumulative curvature) in FEV1 and mortality in patients with CF using UK CF Registry data and found a significant positive association under all TB-WIV definitions. The findings also suggest added predictive value of TB-WIV in FEV1, with cumulative curvature estimated using FPCA giving the best survival predictive performance across different model settings.

A key practical message from this work is that the TB-WIV association parameter (interpreted as the log-hazard ratio in our specification) cannot, in general, be estimated well in irregular and sparse longitudinal settings, even with state-of-the-art semiparametric modelling approaches. As described in Section 4.4 of the paper, the bias patterns are, to some extent, systematic and explainable, and depend on the functional form of TB-WIV and the modelling approach. For FPCA and P-spline-based approaches, the challenge mainly arises from weak identification of curvature patterns in the trajectories, despite generally recovering the underlying mean process very well. For other approaches (SMRE and R-spline), bias can arise further from possible misfit in the mean process.
In practice, if the magnitude of the TB-WIV association is of direct interest, one should be cautious in choosing the modelling approach and in interpreting the estimated effect size. For current curvature, we recommend the P-spline approach in the absence of delayed entry, or the FPCA approach when delayed entry is present; both tend to give estimates close to the true association, though some attenuation may occur. For cumulative curvature, however, we would not treat the estimated magnitude of association as reliable, irrespective of the modelling approach. In any case, we do not recommend the R-spline approach given its stability issues. We expect these implications to extend to the weighted cumulative TB-WIV measure proposed in \citet{wang2024weighted}, with patterns aligning more closely with current or cumulative curvature depending on the spread of the weighting function. In this case, an additional complication is the identification of the weighting function itself, which is particularly challenging under sparse longitudinal sampling. 
Another practical consideration is prior specification. With limited data, a strong prior (e.g., on the basis coefficients) will influence results. In the absence of oracle information, which is most likely in practice, it is reasonable to use weakly informative priors, as considered in this paper. Our experience suggests that, across all three semiparametric approaches, once the relevant priors are made relatively vague, the impact of specific hyperparameter settings is minimal.
Finally, we note that although the TB-WIV association is difficult to estimate reliably and accurately in realistic longitudinal data settings, the utility of incorporating TB-WIV in the joint model remains justified. First, detecting the presence or absence of a TB-WIV association can generally be achieved reliably and is of interest in its own right. Second, we see evidence of added predictive value from TB-WIV, indicating that meaningful information can still be captured, particularly with FPCA and P-spline-based approaches.

There are several directions for further work. Our analysis of the UK CF data indicates the predictive value of TB-WIV, with the resulting JM suggesting improved survival prediction compared with the RB-WIV formulation, as measured by LOOIC. A more systematic investigation of predictive performance gains from including TB-WIV as a predictor, alongside direct comparison with RB-WIV, under dynamic risk prediction using calibration and discrimination metrics in this CF setting, would be of immediate interest. There is also scope to improve modelling and estimation of TB-WIV with denser longitudinal data. Current FPCA estimates eigenfunctions tailored to reconstruction of the mean process, which is not efficient in representing the derivative space, as directions explaining high variability in derivatives can contribute little to the original space. One possible direction is to extend FPCA to make the eigenfunctions more curvature-aware, for example by using a combined objective function defined over both mean and derivative spaces. One may also consider a fully nonparametric approach by employing Gaussian process (GP) priors, with suitable kernel functions, on the trajectories. The three semiparametric approaches studied in this paper are all basis-function-based, and with Gaussian priors on coefficients, they induce a finite-rank GP prior for the subject-specific trajectories, with kernels that are data-driven in FPCA, partly data-driven and rank-restricted in SMRE, and prespecified by the basis-penalty construction in P-splines. The main challenge in moving to a GP formulation lies in computation, and scalable inference algorithms would be required, possibly through variational approaches or other approximation schemes. Another important direction requiring further consideration is to extend the notion of biomarker WIV to more general data scenarios and assess its clinical relevance. One example is discrete-valued markers, which are also common in clinical settings. For multivariate markers, WIV measures that jointly account for marker-specific curvature and cross-marker covariation may also be of interest.

\section*{Author contributions (CRediT)}
Conceptualisation: all authors. Formal Analysis: SC. Funding acquisition: JKB, BDMT. Methodology: SC. Project administration: SC. Software: SC. Supervision: JKB, BDMT. Visualization: SC. Writing –
original draft: SC. Writing – review \& editing: all authors.

\section*{Acknowledgements}
SC was supported by the UK Medical Research Council (MRC) grant “Looking beyond the mean:
what within-person variability can tell us about dementia, cardiovascular disease and cystic fibrosis” (MR/V020595/1) and MRC Unit theme (grant MC\_UU\_00040/02 – Precision Medicine) funding.
JKB was supported through the UK MRC Unit theme (grant MC\_UU\_00040/02 – Precision Medicine) funding. 
MP’s research was supported by the Ulverscroft Vision Research Group (UCL). 
JP was supported by the National Natural Science Foundation of China (Grant Number: 12271047).
BDMT was supported by a George S. Saden Visiting Fellowship funded by the MacMillan Center for International and Area Studies and the Provost’s Office, and by the UK MRC Unit theme (grant MC\_UU\_00040/02 – Precision Medicine) funding. 
For the purpose of open access, the authors have applied a Creative Commons Attribution (CC BY)
licence to any Author Accepted Manuscript version arising from this submission.
We thank the Cystic Fibrosis Epidemiological Network (CF-EpiNet) Strategic Research Centre data
group and Elaine Gunn for contributions to data preparation and cleaning.
The authors declare no conflicts of interest.

\section*{Data availability}
This work used anonymised data from the UK Cystic Fibrosis Registry, which has research ethics approval (Research Ethics Committee reference number 07/Q0104/2). Use of the data was approved by the
Registry Research Committee (data request 426). Data are available following application to the Registry Research Committee (www.cysticfibrosis.org.uk/the-work-we-do/uk-cf-registry/apply-for-data-from-theuk-cf-registry). The R and Stan code used in the simulation studies and case study are available in the GitHub repository https://github.com/sidachen55/TB-WIV.

\bibliographystyle{apalike}
\bibliography{sample}

\clearpage

\setcounter{section}{0}
\setcounter{figure}{0}
\setcounter{table}{0}
\setcounter{equation}{0}
\renewcommand{\thesection}{S\arabic{section}}
\renewcommand{\thefigure}{S\arabic{figure}}
\renewcommand{\thetable}{S\arabic{table}}
\renewcommand{\theequation}{S\arabic{equation}}

\section*{Supplementary materials to the main manuscript}

\begin{figure}[H]
\newlength{\imgrow}
\settoheight{\imgrow}{\includegraphics[width=0.43\textwidth]{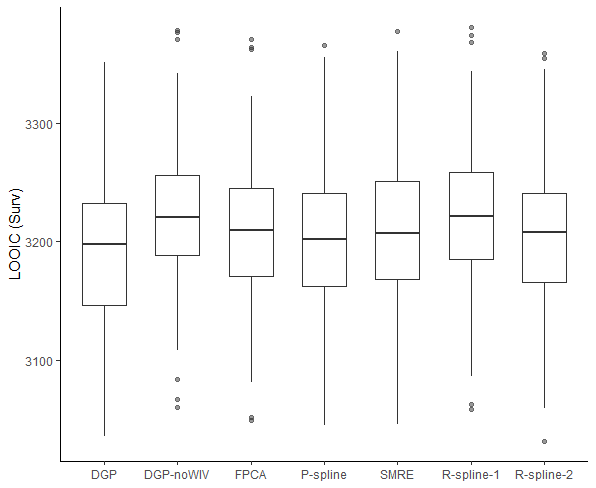}}

\setlength{\tabcolsep}{2pt}
\begin{tabular}{@{}c c c@{}}
  & \textbf{Current curvature} & \textbf{Cumulative curvature} \\
  \raisebox{.5\imgrow}{\rotatebox[origin=c]{90}{\textbf{Case 1}}} &
  \includegraphics[width=0.43\textwidth]{sim_case1_currcurv_looic.png} &
  \includegraphics[width=0.43\textwidth]{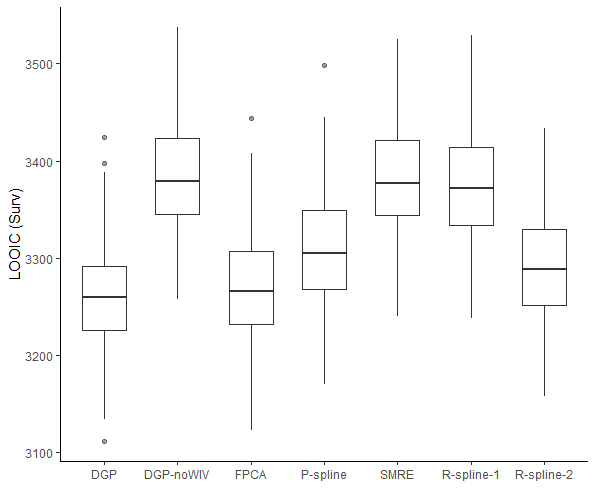} \\
  \raisebox{.5\imgrow}{\rotatebox[origin=c]{90}{\textbf{Case 2}}} &
  \includegraphics[width=0.43\textwidth]{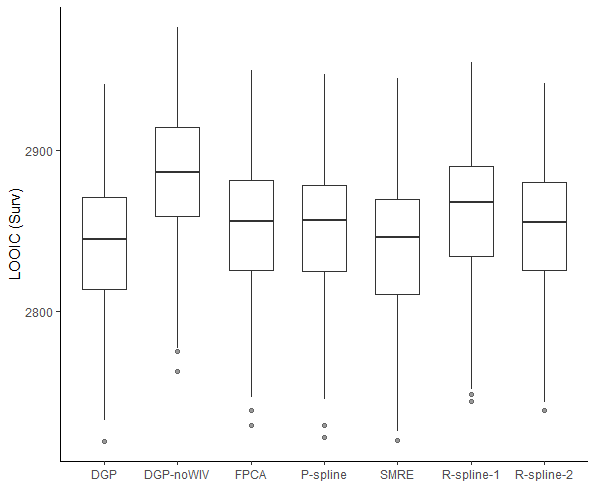} &
  \includegraphics[width=0.43\textwidth]{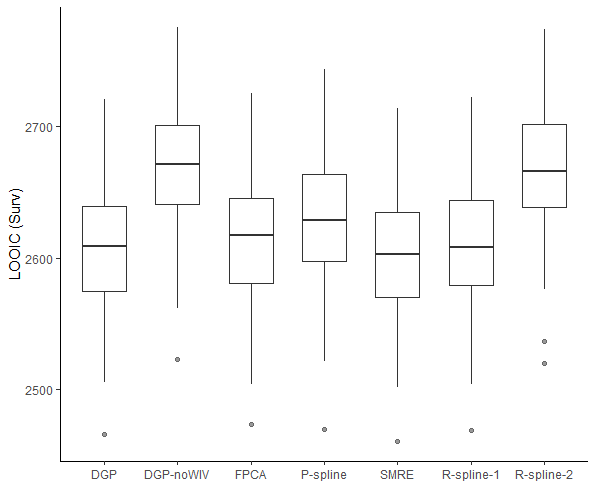} \\
  \raisebox{.5\imgrow}{\rotatebox[origin=c]{90}{\textbf{Case 3}}} &
  \includegraphics[width=0.43\textwidth]{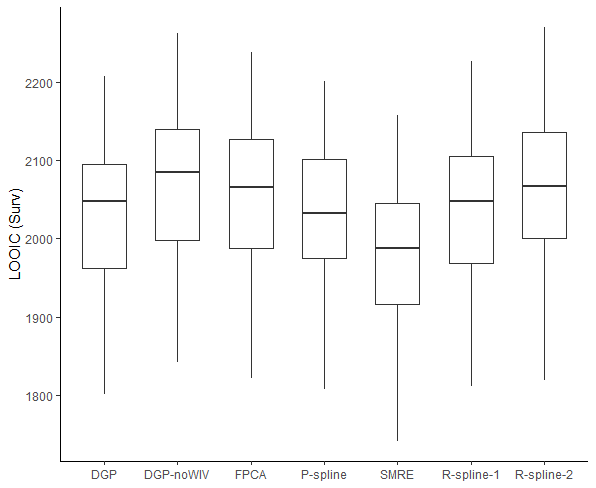} &
  \includegraphics[width=0.43\textwidth]{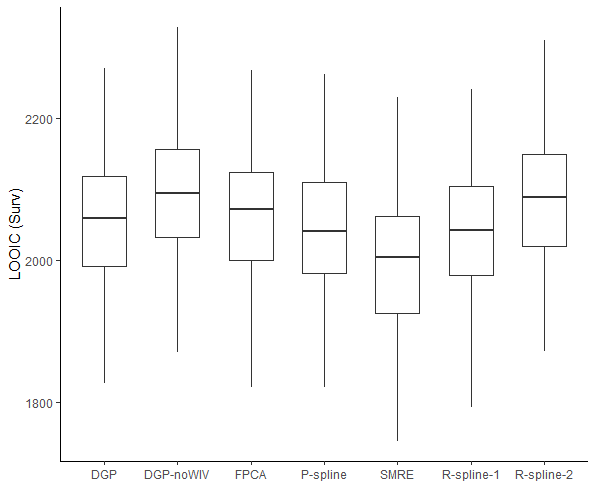} \\
\end{tabular}
\caption{Boxplots of LOOIC for the survival outcome under different modelling approaches: DGP (the true model), DGP-noWIV (DGP without the WIV term), FPCA, P-spline, SMRE, and R-spline-1/2 (as described in Sections 3 and 4 of the main manuscript). The top, middle, and bottom rows correspond to Simulation Cases 1, 2, and 3, respectively. Left panels show results under the current curvature association; right panels show results under the cumulative curvature association. Each panel is based on 100 replicated datasets.}
\end{figure}

\begin{table}[t]
\centering
\caption{Simulation results under Case~3 (null current or cumulative curvature association) comparing DGP (true model), P-spline, FPCA, SMRE, and R-spline approaches (R-spline-1: one interior knot at midpoint; R-spline-2:  three interior knots at equidistant quantiles). Estimated parameters include the fixed effects in the longitudinal submodel ($\beta_1, \beta_2, \beta_3$), those in the survival submodel ($\gamma_1, \gamma_2, \gamma_3$), the association parameter for the current biomarker value ($\alpha_1$), the association for TB-WIV ($\alpha_2$), and the measurement error variance ($\sigma_e^2$). For each parameter, bias is computed from the posterior means across 100 data replications, and coverage probability (CP) is the proportion of times the 95\% credible interval contains the true parameter value across these replications.}
\begingroup
\small
\setlength{\tabcolsep}{3pt}
\renewcommand{\arraystretch}{1.1}
\begin{tabular}{l*{6}{cc}}
\toprule
 & \multicolumn{2}{c}{DGP}
 & \multicolumn{2}{c}{P-spline}
 & \multicolumn{2}{c}{FPCA}
 & \multicolumn{2}{c}{SMRE}
 & \multicolumn{2}{c}{R-spline-1}
 & \multicolumn{2}{c}{R-spline-2} \\
\cmidrule(lr){2-3}\cmidrule(lr){4-5}\cmidrule(lr){6-7}\cmidrule(lr){8-9}\cmidrule(lr){10-11}\cmidrule(lr){12-13}
True value & Bias & CP
     & Bias & CP
     & Bias & CP
     & Bias & CP
     & Bias & CP
     & Bias & CP \\
\midrule
\multicolumn{1}{l}{\textbf{current curvature}}\\
$\beta_{1}=0.25$    & $0.00$ & $94.0$ & $0.00$ & $94.0$ & $-0.01$ & $90.0$ & $0.00$ & $93.0$ & $0.00$ & $94.0$ & $-0.01$ & $83.0$ \\
$\beta_{2}=-0.19$   & $0.00$ & $93.0$ & $0.01$ & $91.0$ & $0.01$  & $93.0$ & $0.01$ & $96.0$ & $0.01$ & $96.0$ & $0.01$  & $80.0$ \\
$\beta_{3}=0.13$    & $0.00$ & $95.0$ & $0.04$ & $91.0$ & $-0.01$ & $86.0$ & $-0.06$& $86.0$ & $0.00$ & $91.0$ & $-0.03$ & $73.0$ \\
$\gamma_{1}=-0.19$  & $-0.01$& $96.0$ & $-0.02$& $97.0$ & $-0.02$ & $96.0$ & $0.00$ & $97.0$ & $-0.02$& $97.0$ & $-0.05$ & $96.0$ \\
$\gamma_{2}=0.02$   & $0.00$ & $95.0$ & $0.00$ & $96.0$ & $0.01$  & $96.0$ & $-0.01$& $94.0$ & $0.00$ & $95.0$ & $0.03$  & $96.0$ \\
$\gamma_{3}=0.04$   & $-0.01$& $93.0$ & $0.00$ & $94.0$ & $-0.01$ & $94.0$ & $0.04$ & $91.0$ & $0.00$ & $93.0$ & $-0.03$ & $93.0$ \\
$\alpha_{1}=-2.95$  & $-0.04$& $96.0$ & $-0.06$& $95.0$ & $0.04$  & $93.0$ & $-0.18$& $75.0$ & $-0.03$& $96.0$ & $0.24$  & $61.0$ \\
$\alpha_{2}=0$      & $-0.16$& $96.0$ & $1.63$ & $89.0$ & $2.44$  & $85.0$ & $-0.91$& $98.0$ & $2.42$ & $97.0$ & $8.00$  & $17.0$ \\
$\sigma_e^2=0.04$   & $0.00$ & $94.0$ & $0.00$ & $97.0$ & $0.00$  & $92.0$ & $0.01$ & $0.0$  & $0.00$ & $0.0$  & $0.00$  & $61.0$ \\

\addlinespace[2pt]
\multicolumn{1}{l}{\textbf{cumulative curvature}}\\
$\beta_{1}=0.24$    & $0.00$ & $93.0$ & $0.00$ & $94.0$ & $-0.01$ & $90.0$ & $0.00$ & $93.0$ & $0.00$ & $93.0$ & $-0.01$ & $82.0$ \\
$\beta_{2}=-0.20$   & $0.00$ & $93.0$ & $0.01$ & $90.0$ & $0.01$  & $94.0$ & $0.01$ & $95.0$ & $0.01$ & $95.0$ & $0.01$  & $80.0$ \\
$\beta_{3}=0.13$    & $0.00$ & $92.0$ & $0.03$ & $90.0$ & $-0.01$ & $87.0$ & $-0.06$& $80.0$ & $-0.01$& $90.0$ & $-0.03$ & $76.0$ \\
$\gamma_{1}=-0.19$  & $-0.02$& $100.0$& $-0.02$& $100.0$& $-0.03$ & $100.0$ & $-0.01$& $96.0$ & $-0.02$& $98.0$ & $-0.05$ & $99.0$ \\
$\gamma_{2}=0.01$   & $0.00$ & $92.0$ & $0.00$ & $91.0$ & $0.00$  & $93.0$ & $-0.02$& $88.0$ & $0.00$ & $91.0$ & $0.03$  & $94.0$ \\
$\gamma_{3}=-0.17$  & $-0.01$& $95.0$ & $-0.02$& $95.0$ & $-0.02$ & $95.0$ & $0.05$ & $94.0$ & $0.00$ & $94.0$ & $-0.26$ & $77.0$ \\
$\alpha_{1}=-3.06$  & $-0.06$& $94.0$ & $-0.09$& $92.0$ & $0.00$  & $96.0$ & $-0.21$& $78.0$ & $-0.06$& $94.0$ & $0.03$  & $97.0$ \\
$\alpha_{2}=0$      & $-0.13$& $94.0$ & $-0.24$& $93.0$ & $0.68$  & $86.0$ & $-0.47$& $88.0$ & $0.57$ & $91.0$ & $2.29$  & $18.0$ \\
$\sigma_e^2=0.04$   & $0.00$ & $95.0$ & $0.00$ & $92.0$ & $0.00$  & $97.0$ & $0.01$ & $0.0$  & $0.00$ & $0.0$  & $0.00$  & $76.0$ \\
\bottomrule
\end{tabular}
\endgroup
\end{table}

\begin{table}[t]
\centering
\caption{Total computational times (hours), including burn-in, for the UK CF data analysis by TB-WIV definition and modelling approach (Section 5 of the main manuscript). Three TB-WIV specifications are considered: current curvature; cumulative curvature; and cumulative curvature with a one-year window.}
\label{tab:realdata_time}
\begingroup
\small
\setlength{\tabcolsep}{6pt}
\renewcommand{\arraystretch}{1.1}
\begin{tabular}{lccc}
\toprule
Joint model setting & FPCA & P-spline & SMRE \\
\midrule
current curvature                      & 6.07 & 9.42 & 6.47 \\
cumulative curvature (one-year window) & 25 & 52.4 &  32.3 \\
cumulative curvature                   & 24.6 & 50.3 & 40.2 \\
\bottomrule
\end{tabular}
\endgroup
\end{table}

\begin{sidewaystable}[!htp]
\centering
\caption{Estimation results obtained from the SMRE-based approach for the UK CF data analysis, as described in Section 5 of the main manuscript. Three TB-WIV specifications are considered: current curvature; cumulative curvature; and cumulative curvature with a one-year window.  Parameters include fixed effects ($\beta_1, \beta_2, \beta_3$ and $\gamma_1, \gamma_2, \gamma_3$), measurement error variance ($\sigma_e^2$), and association parameters ($\alpha_1, \alpha_2$). Posterior means are reported as point estimates (the corresponding posterior distributions are approximately symmetric). The 95\% credible intervals (CIs), shown in brackets, are based on the 2.5th and 97.5th percentiles of the posterior samples.}
\begingroup
\small
\setlength{\tabcolsep}{2.5pt}
\renewcommand{\arraystretch}{1.1}
\begin{tabular}{lccc}
\toprule
Parameter & Current curvature & Cumulative curvature & Cumulative curvature (one-year window) \\
\midrule
\multicolumn{1}{l}{\textbf{Longitudinal}} \\
$\beta_{1}$ -- Diagnosed after year 1     & 0.22 (0.16, 0.28) & 0.21 (0.16, 0.27) & 0.22 (0.16, 0.28) \\
$\beta_{2}$ -- F508del homozygous         & -0.14 (-0.19, -0.09) & -0.14 (-0.20, -0.09) & -0.14 (-0.20, -0.08) \\
$\beta_{3}$ -- Entry age (10y)            & 0.24 (0.19, 0.29) & 0.24 (0.18, 0.29) & 0.24 (0.19, 0.29) \\
$\sigma_e^2$ -- Measurement error variance         & 0.05 (0.04, 0.05) & 0.05 (0.04, 0.05) & 0.05 (0.04, 0.05) \\
\addlinespace[2pt]
\multicolumn{1}{l}{\textbf{Survival}} \\
$\gamma_{1}$ -- Diagnosed after year 1    & -0.18 (-0.33, -0.03) & -0.19 (-0.34, -0.03) & -0.18 (-0.33, -0.03) \\
$\gamma_{2}$ -- F508del homozygous        & 0.01 (-0.13, 0.16) & 0.00 (-0.15, 0.16) & 0.01 (-0.13, 0.15) \\
$\gamma_{3}$ -- Entry age (10y)           & 0.16 (-0.03, 0.36) & 0.11 (-0.09, 0.32) & 0.16 (-0.04, 0.35) \\
$\alpha_{1}$ -- Association (mean)        & -3.02 (-3.21, -2.83) & -3.08 (-3.30, -2.86) & -3.02 (-3.22, -2.82) \\
$\alpha_{2}$ -- Association (TB-WIV)      & 23.05 (12.02, 32.46) & 8.72 (5.50, 12.74) & 23.29 (13.67, 32.76) \\
\bottomrule
\end{tabular}
\endgroup
\end{sidewaystable}

\FloatBarrier
\section{RB-WIV joint model specification (UK CF data application)}
This section provides details on the specification of the joint model with RB-WIV, as considered in Section~5 of the main manuscript. The model structure follows \citet{palma2025bayesian}, with necessary modifications to allow a fair comparison with our proposed model. Specifically, for subject $i=1,\ldots,n$ and visit time $t_{ij}$, $j=1,\ldots,n_i$, the observed longitudinal measurement for FEV1 is modelled using a mixed-effects location--scale framework as
\begin{equation}
    y_i(t) \;=\; m_i(t) + \epsilon_i(t), 
    \qquad \epsilon_i(t)\sim N\!\big(0,\sigma_i^{2}(t)\big),
\end{equation}
with
\begin{equation}
\begin{gathered}
m_i(t) \;=\; w_{L,i}^{\top}\beta^{\mu}_{L} + b^{\mu}_{i0} + \beta^{\mu}_0 + (b^{\mu}_{i1} + \beta^{\mu}_1)t,\\[6pt]
\log \sigma_i(t) \;=\; w_{L,i}^{\top}\beta^{\sigma}_{L} + b^{\sigma}_{i0} + \beta^{\sigma}_0 + \beta^{\sigma}_1 t,
\end{gathered}
\end{equation}
where $w_{L,i}$ is the baseline covariate vector, defined as in the TB-WIV model in Section~5 of the main manuscript, and the vectors $\beta^{\mu}_{L}$ and $\beta^{\sigma}_{L}$ denote the associated coefficients for the location and scale models, respectively. The pairs $(\beta^{\mu}_0,\beta^{\mu}_1)$ and $(b^{\mu}_{i0},b^{\mu}_{i1})$ are fixed and random effects in the location model, while $(\beta^{\sigma}_0,\beta^{\sigma}_1)$ and $b^{\sigma}_{i0}$ are the counterparts in the scale model. The random effects $(b^{\mu}_{i0},b^{\mu}_{i1},b^{\sigma}_{i0})$ are assumed to be jointly Gaussian with mean zero and an unstructured covariance matrix $\Sigma_b$.
Note that, compared to \citet{palma2025bayesian}, we included an additional random slope in modelling the underlying subject-specific trajectory $m_i(t)$ to allow extra flexibility. Preliminary results indicate that the inclusion of a random slope is essential for improving model fit and predictive performance, which is unsurprising given the heterogeneity observed in Figure~1 of the main manuscript.
For the survival submodel, we used a current-value association for both the mean and RB-WIV, as in \citet{palma2025bayesian}:
\begin{equation}
    h_i(t) \;=\; h_0(t)\exp\!\big(w_{S,i}^{\top}\gamma + \alpha_1 m_i(t) + \alpha_2\sigma_i(t)\big),
\end{equation}
where the baseline hazard function $h_0(t)$ is specified as in our TB-WIV model described in Section~5 of the main manuscript, $w_{S,i}=w_{L,i}$ are the baseline covariates included in the longitudinal submodel, and $\gamma$ is the associated coefficient vector. The parameters $\alpha_1$ and $\alpha_2$ represent the association with the biomarker mean level and RB-WIV, respectively.
To complete the Bayesian model specification, we adopted the following set of weakly informative independent priors, following \citet{palma2025bayesian}: for all fixed-effect parameters, including association parameters and spline coefficients, a standard normal prior was assigned. For the covariance matrix $\Sigma_b$, we decomposed it into a vector of standard deviations and a correlation matrix. Each standard deviation was assigned a $\text{half-Cauchy}(0,1)$ prior, and the correlation matrix was assigned an $\text{LKJ}(5)$ prior. Posterior inference was implemented in RStan, with the likelihood function in the survival part adjusted to account for left truncation, using two independent chains with 1000 iterations for burn-in and a further 1000 retained for inference. Convergence diagnostics indicated satisfactory performance.


\end{document}